\documentclass[11pt]{article}
\usepackage{amssymb}

\usepackage{latexsym}
\usepackage{amsmath}
\usepackage{graphicx}
\usepackage{amsthm}
\usepackage{array}
\usepackage{pb-diagram}
\newcommand{\ed}{\end{document}}

\makeindex

\newcommand{\BC}{\mathbb{C}}
\newcommand{\BR}{\mathbb{R}}

\newcommand{\BZ}{\mathbb{Z}}

\newtheorem{lemma}{Lemma}
\newtheorem{theorem}{Theorem}
\newtheorem{corollary}{Corollary}
\newtheorem{remark}{Remark}

\newtheorem{proposition}{Proposition}
\newtheorem{definition}{Definition}

\newcommand{\beq}{\begin{equation}}
\newcommand{\eeq}{\end{equation}}
\newcommand{\sn}{{S^n}}
\newcommand{\bnp}{{B^{n+1}}}
\newcommand{\bnpc}{{{\bar B}^{n+1}}}

\newcommand{\bn}{{\bf n}}
\newcommand{\bx}{{\bf x}}
\newcommand{\by}{{\bf y}}
\newcommand{\br}{{\bf r}}

\newcommand{\cc}{{\cal C}}
\newcommand{\cci}{{\cc^1}}

\newcommand{\lam}{1+\alpha^2+2\alpha(\bn\cdot\bx)}
\newcommand{\tA}{{\tilde A}}
\newcommand{\spo}{Spin}
\newcommand{\spi}{{Spin^1}}
\newcommand{\trg}{{\cal T}}

\newcommand{\tn}{{\cal M}}
\newcommand{\tnp}{{\cal M}_+}
\newcommand{\tno}{{\cal M}_{0+}}
\newcommand{\tnpo}{{\cal M}_{1+}}
\newcommand{\tnpoc}{{\bar \tn}_{1+}}
\newcommand{\tg}{{\cal G}}
\newcommand{\tgp}{{\tg_+}}

\def\vsgap{\vskip 2pt}

\def\vgap{\vskip 10pt}
\DeclareGraphicsExtensions{.jpg,.pdf,.gif,.png,.eps,.bb}

\begin{document}

\title{Quantum fractals on $n$--spheres.\\Clifford Algebra approach.}

\author{Arkadiusz Jadczyk}
\date{\today}
\maketitle
\begin{abstract}
Using the Clifford algebra formalism we extend the quantum jumps
algorithm of the Event Enhanced Quantum Theory (EEQT) to convex
state figures other than those stemming from convex hulls of complex
projective spaces that form the basis for the standard quantum
theory. We study quantum jumps on n-dimensional spheres, jumps that
are induced by symmetric configurations of non-commuting state
monitoring detectors. The detectors cause quantum jumps via
geometrically induced conformal maps (M\"obius transformations) and
realize iterated function systems (IFS) with fractal attractors
located on n-dimensional spheres. We also extend the formalism to
mixed states, represented by ``density matrices" in the standard
formalism, (the n-balls), but such an extension does not lead to new
results, as there is a natural mechanism of purification of states.
As a numerical illustration we study quantum fractals on the circle
(one-dimensional sphere and pentagon), two--sphere (octahedron), and
on three-dimensional sphere (hypercube-tesseract, 24 cell, 600 cell,
and 120 cell). The attractor, and the invariant measure on the
attractor, are approximated by the powers of the Markov operator. In
the appendices we calculate the Radon-Nikodym derivative of the
$SO(n+1)$ invariant measure on $\sn$ under $SO(1,n+1)$
transformations and discuss the Hamilton's ``icossian calculus" as
well as its application to quaternionic realization of the binary
icosahedral group that is at the basis of the 600 cell and its dual,
the 120 cell.

As a by-product of this work we obtain several Clifford algebraic
results, such as a characterization of positive elements in a
Clifford algebra $\cc(n+1)$ as generalized Lorentz ``spin--boosts",
and their action as Moebius transformation on n-sphere,  and a
decomposition of any element of $Spin^+(1,n+1)$ into a spin--boost
and a spin--rotation, including the explicit formula for the
pullback of the $SO(n+1)$ invariant Riemannian metric with respect
to the associated M\"obius transformation.
\end{abstract}
\newpage
\newpage
\section{Introduction}
\begin{quotation} ``The accepted outlook of quantum mechanics (q.m.)
is based entirely on its theory of measurement. Quantitative results
of observations are regarded as the only accessible reality, our
only aim is to predicts them as well as possible from other
observations already made on the same physical system. This pattern
is patently taken over from the positional astronomer, after whose
grand analytical tool (analytical mechanics) q.m. itself has been
modelled. But the laboratory experiment hardly ever follows the
astronomical pattern. The astronomer can do nothing but observe his
objects, while the physicist can interfere with his in many ways,
and does so elaborately. In astronomy the time--order of {\em
states\,} is not only of paramount practical interest (e.g. for
navigation), but it was and is the only method of discovering the
{\em law\,} (technically speaking: a hamiltonian); this he rarely,
if ever, attempts by following a single system in the time
succession of its states, which in themselves are of no interest.
The accepted foundation of q.m. claims to be intimately linked with
experimental science. But actually it is based on a scheme of
measurement which, because it is entirely antiquated, is hardly fit
to describe any relevant experiment that is actually carried out,
but a host of such as are for ever confined to the imagination of
their inventors."\end{quotation} So wrote Ervin Schr\"odinger fifty
years ago \cite{schr55}. Today the standard scheme of q.m. is as
antiquated as it ever was, and provides no answer to the most
fundamental questions such as ``what is time?", and how to describe
{\em events\,} that happen in a single physical system, such as our
Universe.\footnote{Nowadays the defenders of the ``antiquated
scheme" of q.m. go as far as to assign ``crackpot index'' to those
who question this scheme. So, for instance, 10 points (on the scale
of 1--50), are assigned for each claim that quantum mechanics is
fundamentally misguided, and another 10 points for arguing that
while a current well-established theory predicts phenomena
correctly, it doesn't explain ``why" they occur, or fails to provide
a ``mechanism" \cite{baezcrackindex}.} The present paper follows the
line of ideas developed in a series of papers that has led to the
Event Enhanced Quantum Theory (EEQT), as summarized in
\cite{blaja95a}, and recently extended in \cite{jad06a}, but we now
go beyond that framework. While, following von Neumann, we keep the
algebraic structure as one of the most important for the
mathematical formalism of q.m., and we propose to dispose of the
concept of ``observables" and of ``expectation values" at the
fundamental level. We also dispose of the concept of ``time",
understood as a ``continuous parameter", external to the theory. Our
philosophy, concerning ``time" is that of the German social
philosopher Ernest Bloch:
\begin{quotation}``{\em Zeit ist nur dadurch, da\ss{} etwas geschieht und nur dort wo
etwas geschiecht.}\end{quotation} So, time is only  {\em then\,},
when something happens, and only {\em there\,} where something
happens. Therefore the primary concept is that of an {\em event\,},
and of the {\em process\,} - that is a sequence of events. Time, as
a continuous, global variable, comes in only in the limit of a large
number of events. The primary process is that of ``quantum jumps".
It is an irreversible process in an open system, and every system in
which the ``future" is only ``probable", rather than determined, is
necessarily an open system. The mathematical formalism of the
standard quantum theory is based on complex Hilbert spaces and
Jordan algebras of self--adjoint operators. It involves
interpretational axioms for expectation values and eigenvalues of
self--adjoint operators as ``possible results of measurements", yet
it does not provide a framework for {\em defining the
measurements\,} \cite{bell89,bell90}. In view of these
considerations, Gell-Mann would certainly score a high crackpot
index \cite{baezcrackindex} for this statement \cite{gell94}:
\begin{quotation}``Those of us working to construct the modern
interpretation of quantum mechanics aim to bring to an end the era
in which Niels Bohr's remark applies: `If someone says that he can
think about quantum physics without becoming dizzy, that shows only
that he has not understood anything whatever about
it'."\end{quotation} The same can be said about the last paragraph
of Schr\"odingers paper \cite{schr55}, where he wrote
\begin{quotation} ``We are also supposed to admit that the extent of
what is, or might be, observed coincides exactly with what quantum
mechanics is pleased to call observable. I have endeavored to
adumbrate that it does not. And my point is that this is not an
irrelevant issue of philosophical taste; it will compel us to recast
the conceptual scheme of quantum mechanics." \end{quotation} The
need for an open--minded approach is well noted by John A. Wheeler,
who ends his book ``Geons, Black Holes \& Quantum Foam"
\cite{wheeler98} with the following quote from Niels Bohr's friend
Piet Hein:
\begin{quotation}\begin{center}I'd like to know\\
what this whole show\\
is all about\\ before it's out.\end{center}\end{quotation} Alain
Connes and Carlo Rovelli \cite{conro94} proposed to explain the
classical time parameter as arising from the modular automorphism
group of a KMS state on a von Neumann algebra over the field of
complex numbers  $\BC.$\footnote{C.f. also I.~ and G. Bogdanov, {\em
Avant le Big Bang : La cr\'eation du monde\,}, second, revised and
extended edition, (LGF, Paris 2006), where a similar idea, based on
a KMS equilibrium state is discussed in a broader, philosophical
framework} But their philosophy applies, at most, to equilibrium
states, while ``quantum foams" before the Planck era are certainly
far from equilibrium. David Hestenes \cite{hestenes66,hestenes67}
proposed to understand the role of the complex numbers in quantum
theory in terms of the Clifford algebra. This is also our view. L.
Nottale, in his theory of ``scale relativity"  \cite{nottale07}
proposed an alternative idea, where the complex structure arises
from a stochastic differential equation in a fractal space--time. We
think that our approach may serve as a connecting bridge between
fractality, the nontrivial topology of dodecahedral models of
space--time, as discussed by J--P.~Luminet et al. \cite{luminet03}
(cf. also \cite{weeks04}.), and the late thoughts of A. Einstein
\cite[p. 92]{einstein_late}, who wrote:
\begin{quotation}``To be sure, it has been pointed out that the introduction
of a space-time continuum may be considered as contrary to nature in
view of the molecular structure of everything which happens on a
small scale. It is maintained that perhaps the success of the
Heisenberg method points to a purely algebraical method of
description of nature, that is to the elimination of continuous
functions from physics. Then, however, we must also give up, by
principle, the space-time continuum. It is not unimaginable that
human ingenuity will some day find methods which will make it
possible to proceed along such a path. At the present time, however,
such a program looks like an attempt to breathe in empty
space."\end{quotation}

The present paper is a technical one. It fills the empty space with
discrete structures, and it deals with the discrete random aspects
of quantum jumps generated by the algebraic structure of real
Clifford algebras of Euclidean spaces, and of their conformal
extensions. The jumps are generated by M\"obius transformations and
lead to iterated function systems with place dependent
probabilities, thus to fractal patterns on $n$--spheres. Our ideas
are close to those of W. E. Baylis, who also noticed \cite{baylis03}
the similarities between the Clifford algebra scheme and the formal
algebraic structure of q.m. Our results concern the case of the
signature $(1,n+1).$  With some adaptation, the methods and the
ideas developed in the present paper should be also applicable to
the ``hyperbolic quantum formalism", such as developed in recent
papers by A. Khrennikov \cite{khrennikov}.

In Sec. \ref{sec:notation} we introduce our notation, which is kind
of a mixture of that used by Deheuvels \cite{deheuvels} on one hand,
and of Gilbert and Murray \cite{gilbert} on the other. In
Proposition \ref{prop:1} we recall the vector space isomorphism
between the Clifford algebra and the exterior algebra, and in
Proposition \ref{prop:2} we define the trace functional, and list
its properties that are important for applications to quantum
probabilities. In Sec. \ref{sec:vsi} we review the necessary
concepts and results from the monograph by Gilbert and Murray
\cite{gilbert}, and discuss in details the algebra isomorphism
between $\cc^+(1,n+1)$ and the algebra $\BR(2,\cc(n+1)).$ The main
results of this section are given in Theorem \ref{thm:1}.

In Sec. \ref{sec:moebius} we use the Clifford algebra approach to
discus M\"obius transformations of the spheres $S^n,$ as well as
their natural extensions to their interiors $B^{n+1}.$ The key
concept here is that of ``positivity". In Propositions \ref{prop:3}
and \ref{prop:4} we characterize the positive elements of the
$Spin^+$ group, and in Corollary \ref{cor:1} we prove the polar
decomposition of any element of the $Spin^+$ group into a product of
a positive spin--boost and of a unitary spin--rotation. In Theorem
\ref{thm:3} we describe the action of $Spin^+(1,n+1)$ as the
two--fold covering group of the group of M\"obius transformations of
$\sn,$ and give the explicit formula for the action of spin--boosts
on $\sn$ (cf. Eq. (\ref{eq:gxg1})). We also calculate the
Radon-Nikodym derivatives of the transformed surface area and the
volume (cf. Eqs. (\ref{eq:jac}), (\ref{eq:dV})). In subsection
\ref{sec:stereo} we discuss the stereographic projection, and we use
the exponential form of the spin--boosts in order to describe their
(singular) action on $\BR^n$ rather than on $\sn.$

In Sec. \ref{sec:ifs} we discuss iterated function systems (IFS) of
conformal maps and introduce the important concept of the Markov
operator, which is later being used in our numerical simulations
(cf. Sec. \ref{sec:examples}). Proposition \ref{prop:5} of this
section is important in applications to quantum theory. One of the
most important features of the standard, linear, quantum mechanics
is the fact that ``observables" are restricted to bilinear functions
on pure states. Therefore different mixtures of pure states leading
to the same ``density matrix" are claimed to be experimentally
indistinguishable. In our Proposition \ref{prop:5}, and in Corollary
\ref{cor:2}, we show that if the probabilities of the iterated
function systems of M\"obius transformations are given by
geometrical factors derived from the maps themselves (cf. Eqs.
(\ref{eq:gxg1}),(\ref{eq:p12})), and also satisfy the additional
balancing condition (\ref{eq:p11})), then the Markov operator
restricts to the space of functions on $S^n$ given by the trace on
the Clifford algebra, thus leading to a linear Markov semi-group.
Corollary \ref{cor:3} gives the explicit form of the Markov operator
for the case when the IFS of M\"obius transformations is endowed
with geometrical probabilities given by Eq. (\ref{eq:p12}).

Sec. \ref{sec:examples} contains the results of the numerical
simulations of IFS of M\"obius transformations that lead to
``quantum fractals". We study quantum fractals on the circle
(one-dimensional sphere and pentagon), two--sphere (octahedron), and
on three-dimensional sphere (hypercube-tesseract, 24 cell, 600 cell,
and 120 cell). The last section contains the summary and conclusions
and also points out some open problems.

 In the Appendix 1, which is
of independent interest,  we discuss the M\"obius transformation in
terms of the group $SO^+(1,n+1)$ and derive the Radon--Nikodym
derivative formula for a general $SO^+(1,n+1)$ transformation.
Appendix 2 reproduces the original Hamilton's paper of 1856
introducing the ``icossian calculus", while in Appendix 3 we discuss
its application to quaternionic realization of the binary
icosahedral group that is at the basis of 600 cell and its dual, the
120 cell.
\section{Notation\label{sec:notation}}
We will denote by $\BR$ the field of real numbers, and by $\BR^*$
the multiplicative group $\BR\setminus\{0\}.$ Let $V$ be an
$n$--dimensional real vector space endowed with a non--degenerate
quadratic form $Q$ of signature $(r,s),\, r+s=n.$ That is $V$ admits
an orthonormal basis $e_i,$ with $Q(e_1)=\ldots\, =Q(e_r)=1,$
$Q(e_{r+1})=\ldots\, =Q(e_n)=-1.$ Let $\cc=\cc(V,Q)$ the Clifford
algebra of $(V,Q).$ The even and the odd parts of $\cc$ are denoted
as $\cc^+$ and $\cc^-$ respectively. We shall consider $\BR$ and $V$
as vector subspaces of $\cc,$ so that $v^2=Q(v)\in\cc,\, v\in V.$

The principal automorphism of $\cc$ is denoted by $\pi$ and is
determined by $\pi(v)=-v,\, v\in V,$ while the principal
anti--automorphism $\tau,$ denoted also as $\tau(a)=a^\tau ,$ is
determined by $v^\tau=v,\, v\in V. $ Their composition $\nu$ is the
unique anti--automorphism satisfying $\nu (v)=-v$ for all $v\in V.$
We will denote by $\Delta$ the norm function $\Delta:
\cc\longrightarrow \cc,$ defined by \beq \Delta(a)=a^\nu a.\eeq We
recall  that, cf. \cite[5.14--5.16]{gilbert}, if
$\Delta(a),\Delta(b)\in\BR,$ then $\Delta(ab)=\Delta(a)\Delta(b),$
$\Delta(\pi(a))=\Delta(\tau(a))=\Delta(a^\nu)=\Delta(a)$ and, for
all $\lambda\in\BR,$  $\Delta(\lambda a)=\lambda^2\Delta(a).$
Moreover, if $\Delta(a)\in \BR^*,$ then $a$ is invertible, and
$a^{-1}=(1/\Delta(a))a^\nu.$ In particular, if $\Delta(a)\in\BR,$
then also $aa^\nu=\Delta(a).$

We denote by $Spin^+(V,Q)$ the group: \beq Spin^+(V,Q)=\{g\in
\cc^+(V,Q): \Delta(g)=1,\ gVg^{-1}=V\}.\label{eq:spin0}\eeq Every
element $g\in Spin^+(V,Q)$ is a product of an even number of
positive unit vectors (i.e. vectors $v \in v$ such that $Q(u)=+1$)
and an even number of negative unit vectors (i.e. $v \in V$ such
that $Q(u)=-1$) -- cf. \cite[Definition IX.4.C]{deheuvels}. The map
$\sigma: Spin^+(V,Q)\rightarrow SO^+(V,Q),\quad \sigma(g):v\mapsto
gvg^{-1}$ is a two--fold covering homomorphism from $Spin^+(V,Q)$
onto $SO^+(V,Q),$ the connected group of ``proper rotations", that
is orthogonal transformations of $(V,Q)$ of determinant one, which
preserve the orientation of maximal negative subspaces of
$V.$\footnote{The group $Spin^+$ is denoted simply as $Spin$ in Refs
\cite{deheuvels},\cite[2.4.2]{angles}, and as $Spin_0$ in
\cite{gilbert}. The case of $(r,s)=(1,1)$ is special, as in this
case the group $Spin^+$ has two disconnected components -- cf. Ref
\cite[p. 369]{deheuvels}.}

We denote by $\BR(n)$ (resp. $\BR(n,\cc)=\cc\otimes \BR(2)$) the
algebra of $n\times n$ matrices with entries from $\BR$ (resp. from
$\cc$).

\subsection{Vector space isomorphism between the Clifford and the
Grassmann algebra} \label{sec:cg} Let us recall that, as  a vector
space, Clifford algebra is naturally graded and isomorphic to the
exterior algebra. In particular we have the following result :
\begin{proposition}\label{prop:1}Let $e_i,$ $i=1,2,\ldots
,n$ be an orthonormal basis for $V,$ and let $e_I:
I=(i_1,i_2,\ldots,i_p),\, 1\leq i_1<i_2<\ldots <i_p\leq n $ be
defined as the Clifford products $e_I=e_{i_1}e_{i_1}\ldots e_{i_p}$,
with $e_I=1$ for $I=\emptyset .$ Then the set $\{e_I\}$ of $2^n$
vectors in $\cc$ is a linear basis of $\cc,$ the subspaces $\cc_p$
generated by $e_I,$ $I=(i_1,\ldots i_p)$ are independent of the
choice of the orthonormal basis $e_i,$ and $\cc$ is the direct sum
of vector subspaces $\cc_p :$ \beq \cc=\bigoplus_{k=0}^n
\cc_p\label{eq:grad}\eeq Moreover, for each $p=0,\ldots ,n$ the
skew--symmetric map $\alpha_p$ from $V\times V\times\, \ldots\,
\times V$ ($p$ times) to $\cc$ given by:
$$\alpha_p(x_1,x_2,\, \ldots\, ,x_p)=\frac{1}{p!}\sum_\sigma (-1)^{\sigma}x_{\sigma 1}x_{\sigma 2}\, \ldots\, x_{\sigma
p},$$ determines an isomorphism of the vector subspace $\Lambda^p\,
V$ of the exterior algebra $\Lambda V$ onto $\cc_p$ that sends
$e_{i_1}\wedge\,\ldots\, \wedge e_{i_p}\in \Lambda^p\, V$ to
$e_{i_1}\,\ldots\, e_{i_p}\in\cc_p\subset \cc.$
\end{proposition}
\noindent {\bf Proof}: c.f. \cite[Theoreme VIII.10]{deheuvels}$\qed$
\vgap

Of particular interest for us will be the subspace $\cc_0\oplus
\cc_1\subset \cc$ of {\em paravectors\,}. We will denote this
subspace by $V^1$ and endow it with the quadratic form $Q^1$ defined
by \beq Q^1(x^0,v)=(x^0)^2-v^2,\, x^0\in \BR, v\in V.\eeq If $Q$ is
of signature$(r,s),$ then $Q^1$ has signature $(s+1,r).$
\subsection{The trace}
\label{sec:tr} We denote by $\Phi$ the linear functional on $\cc $
assigning to each element $a\in \cc$ its scalar part $\Phi(a)=a_0\in
C_0$ in the decomposition (\ref{eq:grad}). Then the following
proposition holds:
\begin{proposition}\label{prop:2} The functional $\Phi$ has the following properties:
\begin{enumerate}
\item[{\rm (i)}] $\Phi (1)=1,$
\item[{\rm (ii)}] $\Phi(a^\tau)=\Phi (a),$ $\forall a\in \cc ,$
\item[{\rm (iii)}] $\Phi (ab)=\Phi(ba),$ $\forall a,b\in
\cc,$
\item[{\rm (iv)}] $(a,b)\stackrel{df}{=}\Phi(a^\tau b)$ is a nondegenerate,
symmetric, bilinear form on \\ $\cc ,$ that is positive definite if
the original quadratic form on $V$\\ is positive definite. We have
$\Phi(a)=(1,a)=(a,1),\, \forall a\in \cc .$
\item[{\rm (v)}]$(ab,c)=(b,a^\tau c)=(a,cb^\tau ),$ $\forall a,b,c\in
\cc.$
\end{enumerate}
\end{proposition}
\noindent {\bf Proof}: (i) and (ii) follow immediately from the
definition. In order to prove (iii) notice that if $\{e_i\},\,
i=1,\,\ldots\, ,n$ is an orthonormal basis in $V,$ $\{e_I\},
I=\{i_1<\,\ldots\, <i_p\}$ is the corresponding basis in $\cc ,$ and
$a=\sum_I a_Ie^I,\, b=\sum_I b_I e_I$ are the decompositions of $a$
and $b$ in the basis $e_I,$ then $\Phi (ab)=\sum_I a_I b_I \Phi (e_I
e_I)=\Phi(ba).$  From the very definition of the scalar product
$(a,b)$ it follows that $(a,b)=\Phi(a^\tau b)=\Phi((a^\tau
b)^\tau)=\Phi(b^\tau a)=(b,a).$ Moreover, we have $(e_I,e_J)$=0 if
$I\neq J,$ and also $(e_I,e_I)= {e_{i_1}}^2\,\dots\,
{e_{i_p}}^2=(-1)^{s(I)},$ where $s(I)$ is the number of negative
norm square vectors in $I.$ In particular $e_I$ is orthonormal with
respect to the scalar product in $\cc ,$ and so (iv) holds. We have
$(ab,c)=\Phi((ab)^\tau c)=\Phi(b^\tau a^\tau c)=\Phi(a^\tau c b^\tau
)= (a,cb^\tau ),$ which establishes (v).$\qed$\vgap
\noindent\begin{remark}: It is easy to see that
$\Phi(a)=(1/2^n)\mathrm{tr}(L(a)),$ where $L(a)$ is the left
multiplication by $a$ acting on $\cc :$ $L(a)b=ab,$ and the trace is
taken over $\cc ,$ see e.g. \cite[p. 601]{weil1} for a general
discussion. Because of this property $\Phi$ will be called a {\em
trace.}\end{remark}\vsgap

We will call an element $a\in \cc$ {\em positive\,}, which we will
write $a\geq 0,$ if $a=a^\tau$ and $(v,a v)\geq 0$ for all
$v\in\cc.$ Equivalently, $a\geq 0$ if and only if $a$ is of the form
$a=b^\tau b,$ for some $b\in \cc$ (cf. e.g. \cite[7.27]{axler}). If
$a$ is positive and $a\neq 0,$ we will write $a>0.$ If $a\geq 0,$
then, in particular, $\Phi(a)=(1,a 1)\geq 0$ and, if $a>0$ then
$\Phi(a)>0.$
\section{Algebra isomorphism between
$\cc^+(V^1,Q^1)$\\ and $\BR(2,\cc(V,Q))$} \label{sec:vsi} It is a
well known fact (see e.g. Ref. \cite[I.6.13]{gilbert}) that the
algebras $\cc^+(V^1,Q^1)$ and $\cc(V,Q)$ are isomorphic. For the
purpose of the present paper it is useful to have a description of
this isomorphism in some details.\vgap \noindent {\bf Notation}: In
what follows we will use the notation $\cc=\cc(V,Q),$ and
$\cc^1=\cc(V^1,Q^1).$\vgap Let the map $A:\cc\times \cc\rightarrow
\BR(2,\cc)$ be defined by \beq A(a,b)=\left\{
\begin{pmatrix}a&b\\\pi (b)&\pi (a)\end{pmatrix}: a,b\in \cc
\right\}, \label{eq:aab}\eeq and let $\gamma:V^1\rightarrow
\BR(2,\cc)$ be the linear map given by \beq
\gamma(x^0,v)=\begin{pmatrix}0&x^0+v\\x^0-v&0\end{pmatrix}=
\begin{pmatrix}0&x^0+v\\\pi(x^0+v)&0\end{pmatrix}.\label{eq:cmap}\eeq
Then $\gamma$ is evidently the Clifford map,
$\gamma(x^0,v)^2=Q^1(x^0,v)I,$ and therefore it extends to a unique
algebra homomorphism, which we will denote by the same symbol
$\gamma,$ from $\cc^1$ to $\BR(2,\cc).$ We will define now the
following maps, and study their properties: $$ \divide\dgARROWLENGTH
by2
\begin{diagram}
\node{\cc\times\cc}\arrow[2]{e,t}{A}\arrow[2]{s,l}{{\tilde
A}=\,{\tilde \gamma}^{-1}\circ A}\arrow{se,t,-}{A}
\node[2]{Im(A)}\arrow[2]{s}\\
\node{}\node{}\arrow{se}\arrow{ne}\\
\node{\cc^1}\arrow[2]{e,t}{
\gamma}\arrow[2]{se,t}{\psi}\arrow{ne,t,-}{{\tilde \gamma}}
\node[2]{\BR(2,\cc)}\arrow[2]{s,r}{pr_{11}}\\ \\
\node{{\cc^1}^+}\arrow[2]{n}\arrow[2]{e,t}{\psi^+}\node[2]{\cc}
\end{diagram}$$
The map $pr_{11}:\BR(2,\cc)\rightarrow \cc$ assigns to each matrix
in $\BR(2,\cc)$ its top--left entry. For instance
$pr_{11}(A(a,b))=a.$ $Im(A)$ is the set of all matrices of the form
(\ref{eq:aab}). We will not distinguish between the maps
$A:\cc\times\cc\rightarrow \BR(2,R)$ and $A:\cc\times\cc\rightarrow
Im(A),$ which differ only by the canonical inclusion
$Im(A)\rightarrow \BR(2,\cc).$ But we will distinguish between
$\gamma:\cc^1\rightarrow \BR(2,\cc)$ and ${\tilde \gamma}:
\cc^1\rightarrow Im(A).$ The latter map is an algebra isomorphism,
therefore ${\tilde \gamma}^{-1}: Im(A)\rightarrow \cc^1$ is well
defined. The map $\psi$ is defined as $\psi=pr_{11}\circ{\tilde
\gamma},$ and is an algebra homomorphism, and $\psi^+$ is its
restriction to ${\cc^1}^+.$ We will use the notation ${\tilde
A}(a,b)$ for ${\tilde \gamma}^{-1}(A(a,b)).$
\begin{theorem} \label{thm:1}
\begin{enumerate}
\item[{\rm (i)}]
Let us realize the Clifford algebra $\cc(1,-1)$ as the matrix
algebra $\BR(2)$ using the following basis \beq
f_0=\begin{pmatrix}0&1\\-1&0\end{pmatrix},\,
f_1=\begin{pmatrix}0&1\\1&0\end{pmatrix},\,
f_{01}=\begin{pmatrix}-1&0\\0&1\end{pmatrix},\eeq so that we have
\beq f_0^{2k}=1_2,\, f_0^{2k+1}=f_0,\, f_1^{2k}=(-1^k)1_2,\,
f_1^{2k+1}=(-1)^kf_1.\eeq Let $\{e_0\in\BR,e_i\in V,\,
i=1,\ldots,\,n+1\},$ be an orthonormal basis of $V^1.$ Then, in
terms of this basis the map $\gamma:\cc^1\longrightarrow Mat(2,\cc)$
reads:
\begin{eqnarray*} \gamma(1)&=&1_2\otimes1_\cc\\
\gamma(e_0e_{i_1}\ldots\,e_{i_{2k}})&=&(-1)^kf_0\otimes
e_{i_1}\ldots\,
e_{i_{2k}},\\
\gamma(e_0e_{i_1}\ldots\,e_{i_{2k+1}})&=&(-1)^kf_{01}\otimes
e_{i_1}\ldots\,
e_{i_{2k+1}},\\
\gamma(e_{i_1}\ldots\,e_{i_{2k}})&=&(-1)^k1_2\otimes e_{i_1}\ldots\,
e_{i_{2k}},\\
\gamma(e_{i_1}\ldots\,e_{i_{2k+1}})&=&(-1)^kf_1\otimes
e_{i_1}\ldots\,
e_{i_{2k+1}}.\\
\end{eqnarray*}
\item[{\rm (ii)}]
$ker(\psi)={\cc^1}^-,$ and $\psi$ restricts to the algebra
isomorphisms $\psi^+$ from ${\cc^1}^+$ onto $\cc.$ In terms of the
basis we have: \beq \left. \begin{array}{rcl}
\psi^+(1_{\cc^1})&=&1_\cc,\\
\psi^+(e_{i_1}\ldots\,e_{i_{2k}})&=&(-1)^k
e_{i_1}\ldots\,e_{i_{2k}},\\
\psi^+(e_0e_{i_1}\ldots\,e_{i_{2k+1}})&=&(-1)^k e_{i_1}\ldots\,
e_{i_{2k+1}}.\end{array}\right\}\eeq
\item[{\rm (iii)}]
With the notation as above, we have \beq
A(a,b)A(a',b')=A(a'',b''),\, \mbox{where}\quad a''=aa'+b\pi(b'),\,
b''=ab'+b\pi(a').\label{eq:AA}\eeq The principal involution $\pi$
and the principal anti--involution $\tau$ of $\cc^1$ can be
expressed through their corresponding operations in $\cc$ as \beq
\pi(\tA(a,b))=\tA(a,-b),\label{eq:pi}\eeq
\beq\tau(\tA(a,b))=\tA(\nu(a),\tau(b )).\label{eq:tau}\eeq  The even
subalgebra ${\cc^1}^+$ of $\cc^1$ can then be identified with the
set of all $A(a,b),$ with $b=0,$ that is, using the map $pr_{11},$
with $\cc.$
\item[{\rm (iv)}] Denoting by $\Phi^1$ (resp. $\Delta^1),$ and $\Phi$ (resp. $\Delta$) the
trace (resp. norm function) of $\cci$ and $\cc$ respectively, we
have \beq \Phi^1=\Phi \circ  \psi,\eeq \beq \Delta^1({\tilde
a})=\Delta(\psi^+({\tilde a})),\quad \forall\, {\tilde a}\in
\cci,\eeq \beq ({\tilde a},{\tilde b})=(\pi(\psi^+({\tilde
a})),\psi^+({\tilde b})), \quad \forall {\tilde a},{\tilde b}\in
\cci.\eeq

\item[{\rm (v)}] ${\tilde g}\in Spin(V^1,Q^1)$ if and only if
$g=\psi^+({\tilde g})$ satisfies
\begin{enumerate}
\item[{\rm a)}]$\Delta(g)=1,$ and
\item[{\rm b)}] $gV^1g^\tau = V^1.$
\end{enumerate}
\end{enumerate}
\end{theorem}
\noindent{\bf Proof}: \noindent (i) and (ii) follow by a
straightforward calculation.

(iii) By a straightforward matrix multiplication we get from
(\ref{eq:aab}) that \beq A(a,b)A(a',b')=A(a'',b''),\, \mbox{where}\,
a''=aa'+b\pi(b'),\, b''=ab'+b\pi(a').\eeq It follows that the range
(image) of the map $A$ is an algebra and, because it has the right
dimension $2\times dim(V),$ the Clifford map $\gamma$ extends to the
isomorphism of $\cc^1$ onto $Im(A).$ It is also clear that the even
subalgebra of $\cci$ is represented by the matrices $A(a,0),$ while
the odd subspace is represented by matrices $A(0,b).$

It follows from the very definition that $\pi$ and $\tau$ defined by
(\ref{eq:pi}) and (\ref{eq:tau}) are involutions, and that
$\pi(\psi(w))=\psi(-w),$ $\tau(\gamma(w))=\gamma(w)$ for $w\in V^1.$
Therefore we need to show that  $\pi,$ defined by (\ref{eq:pi}), is
an automorphism, and that $\tau,$ defined by (\ref{eq:tau}), is an
anti--automorphism. (Notice that although, by abuse of the notation,
we denote by the same symbol $\pi$ the main automorphisms of $\cc $
and $\cci,$ the meaning is always clear from the context.)

Let $C$ be the matrix\footnote{Cf. also \cite[Xh. VIII.6, p. 310]{deheuvels},
where the matrix $C$ is used to define an anti--involution of the algebra $\BC(2).$}: $C=\left(\begin{array}{cc}1&0\\
0&-1 \end{array}\right),$ then \beq CA(a,b)C^{-1} = A(a,-b),\eeq
therefore the formula (\ref{eq:pi}) defines an involutive
automorphism of $\cci,$ and, since it reverses the signs of vectors,
it defines the principal involution of $\cc^1.$

Proving that $\tau$ is an anti--automorphism of $\cci$ follows by a
straightforward calculation using (\ref{eq:aab}) and the properties
of $\pi$ and $\tau$ on $\cc.$

(iv) follows from (i)--(ii). Finally, (v) follows from (ii) and (iv)
-- (cf. also Ref. \cite[Theorem 6.12]{gilbert}). \qed\vgap From now
on we will assume that $(V,Q)$ is an $(n+1)$-- dimensional Euclidean
space, that is that $Q$ has the signature $(n+1,0).$
\section{M\"obius transformations of $S^n$ and their extensions to
${\bar B}^{n+1}$ \label{sec:moebius}}

\begin{remark}
In chapter 2 of reference \cite{angles}, Pierre Angl\`es gives an
explicit construction of  covering groups of the conformal group of
a standard regular pseudo-euclidean space endowed with a quadratic
form of signature (p,q), together with a geometrical construction of
this conformal group and shows explicitly that this group is
isomorphic to $PO(p+1, q+1),$ by using a wider Clifford algebra
associated with a pseudo-euclidean  regular standard space endowed
with a quadratic form of signature $(p+1,q+1)$ -cf., for example,
\cite[2.4.2.5.2]{angles}. By using his table given in
\cite[2.4.2.4]{angles}, one can characterize the elements of the
conformal group of the sphere $\sn$ stereographically projected onto
$E^n.$ We follow another algebraic method, considering the
particular case of signature $(n,0).$\end{remark}

Let $(V,Q)$ be an $(n+1)$-- dimensional Euclidean space, $n>0.$
Vectors in $V$ we be denoted by bold symbols: $\bx,\by,$ etc. The
bold symbol $\bn$ will be reserved for unit vectors. We will denote
by $\bnp$ the open unit ball $\bnp=\{\bx\in V:\, \bx^2<1\},$ by
$\bnpc$ its closure $\bnpc=\{\bx\in V:\, \bx^2\leq 1\},$ and by
$\sn$ its boundary, the unit sphere $\sn=\{\bn\in V:\, \bn^2=1\}$.
We will denote by $\cc$ the Clifford algebra $\cc(V,Q),$ by $\spo^+$
the group $Spin^+(V,Q),$ and by $\spi^+$ the group
$\psi^+(Spin^+(V^1,Q^1)),$ described by the conditions a) and b) in
Theorem \ref{thm:1}, (v). Following Ref. \cite{gilbert} we define
the Clifford group $\Gamma(V)$ as $$
\Gamma(V)=\{w_1\ldots\,w_k:w_j\in V^1,\quad \Delta(w_j)\neq 0\}.$$
It is evident that this group is closed under $\pi,\tau,\nu.$

We will describe the action of $\spi^+$ on the unit sphere $S^n,$
and on its interior $\bnp.$ As the main tool we will use the special
class of elements of $\cc,$ that are called {\em transformers\, }.
\subsection{Transformers}
Following Gilbert and Murray \cite[5.21]{gilbert} we define a
transformer to be any element $a$ of $\cc$ with the property that
for every element $w\in V^1$ there exists another $w'\in V^1$ such
that \beq aw=w'\pi(a).\label{eq:aw}\eeq The set $\trg$ of all
transformers is a multiplicative semigroup. Moreover we have the
following important result proven in \cite[5.24--5.29]{gilbert}:
\begin{theorem} The set of all transformers $\trg$ is closed under
the principal automorphism $\pi.$ Moreover, for every $a\in\trg,$
$\Delta(a)\in\BR,$ and if $\Delta(a)\neq 0,$ then also
$a^\tau\in\trg.$ The set of all invertible transformers coincides
with Clifford group $\Gamma(V).$
\qed\label{thm:gilbert}\end{theorem}
\begin{lemma}
If $a$ is an invertible transformer, then for every $w\in V^1$ we
have \beq \sigma_a(w)\stackrel{df}{=}awa^\tau\in
V^1.\label{eq:sigma}\eeq
\end{lemma}
\noindent{\bf Proof}: We first notice that $\tau(w)=w,\,\forall w\in
V^1.$ Applying $\tau$ to both sides of the defining equation
(\ref{eq:aw}) we get $wa^\tau=a^\nu w'.$ Multiplying by $a$ from the
left, we get $awa^\tau=aa^\nu w'.$ But since
$\Delta(a)=\Delta(\pi(a))=a^\nu a=aa^\nu\in \BR, $ we get
$awa^\tau=\Delta(a)w'\in V^1.$ \qed\vgap Motivated by the above
lemma we define the subsets $\tn,\tnp\in\cc$ as follows:

\begin{eqnarray}
\tn&=&\{a\in\cc:\, aV^1a^\tau\subset
V^1\}.\\
\tnp&=&\{a\in\tn:\, a>0,\, \Phi(a)=1\} \label{eq:tn}\end{eqnarray}

\begin{definition}We define the following important subsets of $\trg$ and of $\tnp$:
\begin{eqnarray}
\tg&=&\{a\in\trg:\, \Delta(a)=1\},\\
\tg_R&=&\{a\in\tg:\, aa^\tau=1\},\\
\tgp&=&\{a\in\tg:\, a\geq 0\},
\end{eqnarray}
\begin{eqnarray}
\tnpo&=&\{a\in\tnp: \Delta(a)>0\},\\
\tnpoc&=&\{a\in\tnp: \Delta(a)\geq 0\},\\
\tno&=&\{a\in\tnp: \Delta(a)=0\}.
\end{eqnarray}
\end{definition}
\noindent Notice that, by the Theorem \ref{thm:gilbert}, $\tg$ is
invariant under both $\pi$ and $\tau.$ It is sometimes denoted as
$Spoin_0(V),$ and the map $\sigma$ (cf. \ref{eq:sigma}) is a
two--fold covering homomorphism: $\tg\mapsto SO^+(V^1,Q^1)$ -- c.f.
\cite[6.12]{gilbert}. Thus $\tg$ is nothing but $\spi^+ .$ $\tg_R$
leaves the subspace $V\subset V^1$ invariant and $\sigma,$ when
restricted to $\tg_R,$ is a two--fold covering homomorphism of
$SO(V,Q).$ The elements of $\tgp,$ that will be studied in our paper
in some details, will be called {\em spin--boosts\,}. $\tn$ is a
multiplicative semigroup, and $\tg\subset \tn.$ We will show that
$\tno$ is naturally isomorphic to the unit sphere $S^n,$ while
$\tnpo$ (resp. $\tnpoc$) corresponds to the open unit ball $B^{n+1}$
(resp its closure ${\bar B}^{n+1}$).
\begin{lemma} \label{lem:2}
Let $a\in V^1,\, a\neq 0,1,\,\, \Phi(a)=1,\, \Delta(a)\geq 0.$ Then
$a>0,$ and $a$ is of the form $a=1+\alpha\bn,$ $0<\alpha\leq 1,$
$\bn\in V,\, \bn^2=1.$ If $\Delta(a)>0,$ then \beq
\sqrt{a}=\frac{1}{\sqrt{1+\epsilon^2}}(1+\epsilon\bn),\quad
\mathrm{where}\quad \epsilon=\frac{1-\sqrt{1-\alpha^2}}{\alpha}.\eeq
If $\Delta(a)=0,$ then $\alpha=1,$ $a=1+\bn,$ and
$\sqrt{a}=\frac{1}{\sqrt{2}}a.$\end{lemma}
\noindent{\bf Proof}: Since $a\in V^1$, $a=x^0+\bx,$ $x^0\in\BR,$
$\bx\in V.$ Since $\Phi(a)=x^0,$ and $\Delta(a)=(x^0)^2-\bx^2,$ it
follows that $x^0=1,$ $\bx^2\leq 1.$ Let us write $a$ as
$a=1+\alpha\bn,\, 0<\alpha\leq 1,\, \bn^2=1.$ Consider first the
case of $\Delta(a)=1-\alpha^2>0,$ i.e. $\alpha<1.$ Let
$b\stackrel{df}{=}\frac{1}{\sqrt{1+\epsilon^2}}(1+\epsilon\bn),\quad
\mathrm{where}\quad \epsilon=\frac{1-\sqrt{1-\alpha^2}}{\alpha}.$
Then $b^\tau=b,$ and, by simple algebra, we get $0<\epsilon<1,$
$b^\tau b=b^2=a.$ Thus $a>0.$ But now $b$ has the same form as $a$
(up to a positive multiplicative factor), therefore also $b>0.$ Then
$b=\sqrt{a}$ follows from the uniqueness of a positive square root
of a positive element. If $\Delta(a)=0,$ i.e. $a=1+\bn,$ then $a^2=
(1+\bn)^2=1+2\bn+\bn^2=2a,$ and therefore $a>0,$ and
$\sqrt{a}=a/\sqrt{2}.$ \qed\vgap
\noindent The following proposition characterizes explicitly the
sets $ \tno,\tnpo,\tnpoc.$
\begin{proposition}
Let $P: V\supset \bnpc \rightarrow V^1\subset\cc$ be the map \beq
P(\bx)=1+\bx,\quad \bx\in\bnpc .\eeq Then $P$ is a bijection
$P:\bnpc\rightarrow{\bar \tn}^+_1,$  $P(\sn )=\tno,$ and $P(\bnp
)=\tn^+_1.$ \label{prop:3}
\end{proposition} \noindent{\bf Proof}: If
$\bx=0,$ then $P(\bx)=1,$ which is evidently in ${\bar \tn}^+_1.$
Let us therefore assume $0<\bx^2\leq 1.$ With $a=P(\bx),$ we have
$a=a^\tau,$ $\Phi(a)=1,$ $\Delta(a)=1-\bx^2\geq 0,$ therefore, by
Lemma \ref{lem:2}, $a>0.$ Moreover, by a simple calculation, we find
that if $w=y^0+\by\in V^1,$ then \beq
awa=y^0(1+\bx^2)+2(\bx\cdot\by)+(1-\bx^2)\by+2(y^0+(\bx\cdot\by))\bx\in
V^1.\eeq Therefore $a\in {\bar \tn}^+_1.$ To show that $P$ is a
surjection onto $\tnpoc,$ let $a$ be an arbitrary element in $\tnpoc
.$ Then $a^2=a(1+{\bf 0})a$ must be in $V^1.$ Let us therefore write
$a^2=y^0+\by.$ Now $y^0=\Phi(a^2)>0,$ and
$\Delta(a^2)=\Delta(a)^2\geq 0,$ Therefore, we can write
$a^2=y^0(1+\alpha\bn ),$ $\alpha\leq 1.$ Then it follows from Lemma
\ref{lem:2} that $a^2$ has a square root in $V^1$ and, because of
the uniqueness of the square root, $a$ itself must be in $V^1.$ But,
since $\Phi(a)=1,$ and $\Delta(a)\geq 0,$ it follows that $a=1+\bx,$
$\bx^2\leq 1.$ This shows that $P$ is a bijection. The remaining
statements follow from $\Delta(P(\bx))= \Delta(1+\bx)=1-\bx^2.$
\qed\vgap The following proposition and its corollary describe the
set of spin--boosts $\tgp,$ and the Iwasawa--type decomposition of
$\tg.$
\begin{proposition}\label{prop:4}
$m\in \tg_+$
 if and only if $m$ is of the form
\beq m=\frac{1+\alpha\bn}{\sqrt{1-\alpha^2}},\quad \bn\in\sn,\quad
0\leq\alpha<1.\label{eq:ai}\eeq An equivalent form is that of \beq
m=\exp\left(\frac{\eta}{2}\bn\right),\quad \alpha=\tanh (\eta/2),\,
\eta > 0.\label{eq:exp}\eeq
\end{proposition}
\noindent{\bf Proof}: The sufficient condition: With $m,$ $\bn,$ and
$\alpha$ as in (\ref{eq:ai}), it follows from the Proposition
\ref{prop:3} that $1+\alpha\bn>0.$ On the other hand $\Delta(m)=1,$
thus $m\in\tgp.$ On the other hand, since $\bn^2=1,$ is easy to
calculate the exponential in (\ref{eq:exp}), the result being: \beq
\exp\left(\frac{\eta}{2}\bn\right)=\cosh(\eta/2)+
\sinh(\eta/2)\bn.\label{eq:exp1}\eeq It is then easy to see that by
setting $\alpha=\tanh (\eta/2),\, \eta > 0,$ we recover
(\ref{eq:ai}).\\
The necessary condition. We can assume that $m\neq 1.$ Suppose
$m\in\tgp,$ then $\Phi(m)>0,$ and thus $m/\Phi(m)\in \tn^+_1.$ It
follows from the Lemma \ref{lem:pos} that $m$ is proportional to
$1+\alpha\bn,\quad 0<\alpha<1,\, \bn^2=1.$ Then, from $\Delta(m)=1$
it follows that the proportionality coefficient is
$1/\sqrt{1-\alpha^2}.$ \qed
\begin{corollary}\label{cor:1}
$\tg=\tgp \tg_R.$ Every element $g\in\tg$ has a unique decomposition
into the product \beq g=mu,\quad m\in\tgp,\, u\in
\tg_R.\label{eq:dec}\eeq
\end{corollary}
\noindent{\bf Proof}: Let $g\in\tg.$ If $g=1,$ then there is nothing
to prove, as we take $m=1, u=1.$ Let us therefore assume $g\neq 1.$
Using the Polar Decomposition Theorem (cf. e.g. \cite[p.
153]{axler}), $g$ can be written, in a unique way, as $g=mu,$ where
$m^2=gg^\tau>0$, and $uu^\tau=u^\tau u=1.$ We need to show that
$m\in\tgp$ and $u\in\tg_R.$ Now, since $\tg$ is invariant under
$\tau,$ it follows that $m^2=gg^\tau\in \tgp.$ Therefore, by (i),
$m^2$ can be written as $m^2=\exp(\eta\bn/2)$ and, from the
uniqueness of the square root, $m=\exp(\eta\bn/4).$ Therefore
$m\in\tgp.$ It follows that $u=m^{-1}g\in\tg, $ and so $u\in\tg_R.$
\qed
\begin{remark} The decomposition given in (\ref{eq:dec}) corresponds to
the well known decomposition of Lorentz transformations into
``boosts" and ``space rotations." The special case of $n=2,$ and
$SO^+(1,3),$ though not at the Clifford algebra level, is treated in
details in Ref. \cite{moretti}\end{remark} Let us now describe the
action of the group $\tg$ on $\bnpc.$ We will need the following
lemma, which is the result of a simple, though somewhat lengthy,
calculation in the Clifford algebra $\cc.$
\begin{lemma}
If $m=P(\alpha\bn)/\sqrt{1-\alpha^2}\in\tgp,$ then for all $\bx\in
\bnpc$ we have \beq
P(\alpha\bn)(1+\bx)P(\alpha\bn)=(1+\alpha^2+2\alpha\,(\bn\cdot\bx))(1+\bx'),\label{eq:qaq}\eeq
\beq m(1+\bx)m=\frac{\lam }{1-\alpha^2}\, (1+{\bx}^\prime
),\label{eq:gxg}\eeq where \beq {\bx}^\prime=\frac{(1-\alpha^2)\bx
+2\alpha(1+\alpha\,
(\bn\cdot\bx))\bn}{1+\alpha^2+2\alpha\,(\bn\cdot\bx)}.\label{eq:gxg1}\eeq
\end{lemma}\qed\vgap
Before stating the next theorem let us notice that if $\bx^2\leq 1,$
then $P(\bx )=1+\bx
>0.$ If $g\in\tg,$ then also $gP(\bx)g^\tau>0$ and, therefore,
$\Phi(gP(\bx)g^\tau)>0.$ Since $\tg\subset\tn,$ and since $\tn$ is a
multiplicative semigroup, it follows that $gP(\bx)g^\tau\in\tn,$ and
therefore $gP(\bx)g^\tau/\Phi(gP(\bx)g^\tau)\in\tnpoc.$

We also recall the definition of a conformal transformation (see
e.g. Ref. \cite[Ch. 3.7]{goldberg}
\begin{definition}
A diffeomorphism $\phi$ of a Riemannian manifold $(M,G)$ is called a
conformal transformation if there is a function $\rho>0$ on $M$ such
that
$$(\phi^*G)_{\alpha\beta}=\rho^2 G_{\alpha\beta}.$$
If $n=\dim (M)\geq 3$ then the group of conformal transformations of
$M$ is a Lie group of $dim\leq \frac{(n+1)(n+2)}{2},$ and for the
spheres $\sn,$ that are of particular interest in our paper, the
upper limit is reached - cf. e.g. \cite[Note 11, p. 309]{kobayashi}
and also references in \cite[Ch. 2]{angles}.
\label{def:2}\end{definition}
\begin{remark}: The case
of $n=2$ is exceptional, as in this case every complex analytic
transformation of the complex plane generates a conformal
transformation on the Riemann sphere. In this case it is better to
deal with the subgroup of all conformal transformations of $S^2,$
called ``M\"obius transformations." These are the transformations of
$S^n$ that preserve cross--ratios
$$\frac{d(u,x)d(v,y)}{d(u,v)d(x,y)},\qquad u,v,x,y\in S^n,$$
where $d$ is the natural distance on $S^n.$ More information about
various, equivalent definitions and properties of M\"obius
transformations of$S^n$ and of $B^{n+1}$ can be found, for example,
in Refs. \cite[Ch. 4]{ratcliffe} and \cite[Ch.
2]{angles}.\label{rem:rat2}
\end{remark}
\begin{theorem}\label{thm:3}
\begin{enumerate}
\item[{\rm (i)}]
Let for each $g\in\tg,$ let $\phi_g:\bnpc\rightarrow\bnpc$ be
defined by \beq
\phi_g(\bx)=P^{-1}\left(\frac{\sigma_g(P(\bx))}{\Phi(\sigma_g(P(\bx)))}\right)\label{eq:phi}.\eeq
Then $g\mapsto\phi_g$ is a homomorphism from $\tg$ onto a group of
transformations of $\bnpc.$
\item[{\rm (ii)}]
If $m\in\tgp$ is written as in (\ref{eq:ai}):
$m=(1+\alpha\bn)/\sqrt{1-\alpha^2},$ then the M\"obius
transformation $\phi_m$ is explicitly given by the formula: \beq
\phi_m(\bx)=\frac{(1-\alpha^2)\bx +2\alpha(1+\alpha\,
(\bn\cdot\bx))\bn}{1+\alpha^2+2\alpha\,(\bn\cdot\bx)},\quad
\bx\in\bnpc.\label{eq:phim1}\eeq
\item[{\rm (iii)}] When restricted to the unit sphere $\sn,$ $\phi$ is a two--fold
covering homomorphism from $\tg$ onto the group of M\"obius
transformations of $\sn.$
\item[{\rm (iv)}] For $m=(1+\alpha\bn)/\sqrt(1-\alpha^2)\in\tgp,$ the map $\phi_{m}:
S^n\ni\bx\mapsto \bx^\prime\in S^n,$ given by (\ref{eq:gxg1}), is
conformal with the conformal factor \beq
\rho=\frac{(1-\alpha^2)}{(1+\alpha^2+2\alpha(\bn\cdot\bx))}.\label{eq:rho}\eeq
That is, if $G=(G_{\alpha\beta})$ is the natural Riemannian metric
on the unit sphere then \beq (\phi_m^*
G)_{\alpha\beta}=\frac{(1-\alpha^2)^2}{(1+\alpha^2+2\alpha(\bn\cdot\bx))^2}\,G_{\alpha\beta}.\label{eq:metric}\eeq
Thus $\phi_m$ does not, in general, preserve the canonical,
$SO(V)$--invariant, volume form $dS$ of $S^n.$ Denoting by $dS'$ the
pullback\footnote{Let us recall that if $\phi:M\longrightarrow N$ is
a $C^1$ map between differentiable manifolds $M$ and $N,$ and if
$\omega$ is a k--form on $N,$ then its pullback $\phi^*(\omega )$ is
the k--form on $M$ defined by $\phi^*(\omega)({\xi_1}_p,\ldots
,{\xi_k}_p)=\omega(d\phi_p({\xi_1}_p),\ldots ,d\phi_p({\xi_k}_p))$
for all ${\xi_1}_p,\ldots ,{\xi_k}_p\in T_p(M),\, p\in M$ where
$d\phi_p: T_pM\longrightarrow T_{\phi(p)}N$ is the derivative of
$\phi$ at $p$. For a  composition of maps we have $(\phi\circ
\psi)^*=\psi^*\circ \phi^*$ -- cf. e.g. \cite[Ch.
XVI.20]{dieudonne}.} $\phi_{m}^*(dS)$ of $dS$ by $\phi_{m},$ for
every $\bx\in S^n$ we have: \beq \frac{dS^\prime}{dS}(\bx
)=\left(\frac{1-\alpha^2}{\lam}\right)^n.\label{eq:jac}\eeq If the
map (\ref{eq:gxg1}) is applied to the ball $B^{(n+1)}$ (rather than
to its boundary $S^n$), and if $dV$ denotes the standard Euclidean
volume form of $V^1,$ then \beq
\frac{dV^\prime}{dV}=\left(\frac{1-\alpha^2}{\lam}\right)^{n+2}.\label{eq:dV}\eeq
\end{enumerate}
\end{theorem}
\begin{remark}
It is easy to see that our definition of conformal (M\"obius)
transformations of $\sn$ is equivalent to one given by Pierre
Angl\`es in Ref. \cite[2.4.1,2.4.2.1]{angles}. In particular $\tno$
can be identified with $P(Q^1-\{0\})$ in the notation of Ref.
\cite{angles}. But we do not need the stereographic projection that
distinguishes the vector $e_{n+1}\in V.$
\end{remark}
\begin{remark} The transformations $\phi_g:\bnpc\rightarrow
\bnpc,$ defined in (\ref{eq:phi}) are also called Poincar\'e
extensions of those restricted to $\sn$ -- cf. \cite[[Ch. 4.4,
4.5]{ratcliffe}.\end{remark} \noindent{\bf Proof}: (i)
That $\phi_g$ is a group homomorphism follows directly from the
defining formula. In order to  show that each $\phi_g$ maps $\sn$
onto $\sn,$ we first notice that from $\Delta(P(\bx))=1-\bx^2,$ it
follows that $\bx\in\sn$ if and only if $\Delta(P(\bx))=0.$ If
$\Delta(P(\bx))=0,$ then, since $\Delta(g)=\Delta(g^\tau)=1,$ also
$\Delta(gP(\bx)g^\tau)=\Delta(g)^2\Delta(P(\bx))=\Delta(P(\bx))=0,$
thus $\phi_g(S^n)\subseteqq S^n.$ In fact, since
$g^{-1}=g^\nu\in\tg,$ we have that $\phi_g(S^n)=S^n.$

\noindent (ii) Follows from (\ref{eq:gxg}).

\noindent (iii)Let us show that $\phi$ so restricted to $\sn$ is has
kernel $\BZ_2.$ We first notice that if $g\in ker \phi$ then
$g^\tau\in ker \phi.$ Indeed, from the very definition of $\phi$ it
follows that $g\in\ker\phi$ if and only if $g(1+\bn)g^\tau$ is
proportional to $1+\bn$ for all $\bn\in\sn:$
$$g(1+\bn)g^\tau=\lambda(1+\bn).$$ By applying $\pi$ to both sides
of this equation, we get $$\pi(g)(1-\bn)g^\nu=\lambda(1-\bn).$$ Now,
multiplying by $g^\tau$ from the left, and by $g$ from the right,
and taking into account the fact that $\Delta(g)=g^\nu
g=\Delta(g^\tau)=g^\tau\pi(g)=1,$ we find
$g^\tau(1-\bn)g=(1/\lambda)(1-\bn),$ and, since $\bn\in\sn$ is
arbitrary, $g^\tau\in ker \phi.$ Now, assuming that $g\in ker \phi,$
let $g=mu$ be the decomposition of $g$ into a spin--boost $m\in\tgp$
and a rotation $u\in\tg_R.$ Then $g^\tau=u^\tau m\in ker \phi,$ and,
since the kernel of a group homomorphism is a group, we get
$m^2=gg^\tau\in ker \phi,$ i.e. $\phi_{m^2}(\bx)=\bx,\quad
\bx\in\sn.$ Let us write $m^2$ as
$m^2=(1+\alpha\bn)/\sqrt{1-\alpha^2}$ and, since we have assumed
that $dim(V)\geq 2,$ we can choose for $\bx$ a unit vector in $V,$
orthogonal to $\bn.$ Then, from (\ref{eq:gxg1}) we get
$$\bx=\phi_{m^2}(\bx)=\frac{(1-\alpha)^2\bx+2\alpha\bn}{1+\alpha^2},$$
which is possible only for $\alpha=0,$ i.e. if $m^2=1.$ But then,
from the uniqueness of the square root, $m=1,$ and so $g=u.$ Now,
$u(1+\bx)u^\tau=(1+\bx)$ implies $u\bx u^\tau =\bx,$ which extends,
by simple scaling to all $\bx\in V.$ Since
$u^\tau\pi(u)=\Delta(u^\tau)=\Delta(u)=1,$ the last equation can be
rewritten as $u\bx=\bx\pi(u),$ and it follows from \cite[Lemma
5.25]{gilbert} that $u\in\BR.$ Then, since $\Delta(u)=1,$ we get
$u^2=1,$ so that $u=\pm 1.$ That the homomorphism $\phi$ is
surjective, as its image is a connected Lie group of conformal
transformations of dimension equal to that of $Spin^+(1,n+1),$ that
is $(n+2)(n+1)/2$ - cf. Definition \ref{def:2} and Remark
\ref{rem:rat2}.

\noindent (iv) Let us endow $V$ with an orthonormal basis
$e_1,\ldots\, ,e_{n+1}$, and the corresponding coordinates
$x^1,\ldots\, ,x^{n+1}.$ Let $G=(G_{ij}=\delta_{ij})$ be the natural
Riemannian metric in $V.$ From (\ref{eq:gxg1}) it is then easy to
compute $G^*_{ij}=(\phi^*_m G)_{ij}=\frac{\partial x'^k}{\partial
x^i}\frac{\partial x'^l}{\partial x^j}\delta_{kl}.$ The result is
\beq G^*_{ij}=\rho^2\left(
\delta_{ij}+\frac{4\alpha^2(\bx^2-1)}{f^2}\,n_in_j-\frac{2\alpha}{f}(n_ix_j+n_jx_i)\right).\eeq
where \beq f=1+\alpha^2+2\alpha(\bn\cdot\bx),\qquad
\rho=\frac{1-\alpha^2}{f}.\label{eq:rho2}\eeq If ${\bf v}=(v^i)$ and
${\bf w}=(w^i)$ are vectors
 tangent to $\sn,$
so that $({\bf v}\cdot\bn)=({\bf w}\cdot\bn)=0$ then, when computing
$G^*_{ij}v^iw^j,$ the two last terms vanish, and we obtain
$G^*_{ij}v^iw^j=\rho^2 G_{ij}v^iw^j,$ which proves
(\ref{eq:metric}). (\ref{eq:jac}) follows immediately from
(\ref{eq:metric}). It is also easy to calculate the determinant of
the matrix $G^*.$ It has eigenvalue equal to $\rho^2$ on the whole
$(n-1)$-- dimensional subspace orthogonal to $\bn$ and $\bx,$ while
the product of its two eigenvalues in the subspace spanned by $\bn$
and $\bx$ is equal to $\rho^4.$ So the determinant is
$\rho^{2(n+1)},$ and the square root of the determinant is
$\rho^{n+2},$ which proves (\ref{eq:dV}).\footnote{The same way one
gets (\ref{eq:dV}) also for $\bx\| \bn.$ An alternative method of
proving (\ref{eq:jac}) and (\ref{eq:dV}), using $(n+1$)--
dimensional polar coordinates can be found in a previous version of
this paper, available as an arxiv preprint \cite{jadv2}}\qed
\subsection{Stereographic projection}\label{sec:stereo}
In order to get a better insight into the geometrical nature of our
transformations, and also to understand why, in (\ref{eq:exp}),
following Ref. \cite{angles}, we have used $\eta/2,$ rather than
just $\eta$ as the parameter of the exponential, it is instructive
to discuss the action of our transformations on the stereographic
projection of the sphere $S^n.$ As before, we fix the vector $\bn\in
S^n,$ and let $s_\bn$ be the stereographic projection from $S^n$
onto the hyperplane through the origin of $V,$ orthogonal to $\bn,$
with the origin at $\bn.$ Explicitly, we have \beq s_\bn
(\bx)=\frac{\bx-(\bn\cdot\bx)\bn}{1-(\bn\cdot\bx)},\quad \bx\in
S^n.\eeq Indeed, the vector $s_\bn(\bx)$ is on the straight line
connecting $\bn$ and $\bx,$ and is orthogonal to $\bn,$ which two
properties uniquely characterize the stereographic projection. Let
us recall now the action of $\phi_m$ on $S^n.$ From the formula
(\ref{eq:gxg1}) we have:

\beq {\bx}^\prime=\frac{(1-\alpha^2)\bx +2\alpha(1+\alpha\,
(\bn\cdot\bx))\bn}{1+\alpha^2+2\alpha\,(\bn\cdot\bx)}.\label{eq:agxg1}\eeq
Let us compare now $s_\bn(\bx')$ with $s_\bn(x).$ By a
straightforward calculation we obtain: \beq
(\bn\cdot\bx')=\frac{2\alpha+(1+\alpha^2)\bn\cdot\bx}{1+\alpha^2+2\alpha(\bn\cdot\bx)},\eeq
\beq
1-(\bn\cdot\bx')=\frac{(1-\alpha)^2(1-(\bn\cdot\bx))}{1+\alpha^2+2\alpha(\bn\cdot\bx)},\eeq
\beq
\bx'-(\bn\cdot\bx')\bn=\frac{(1-\alpha^2)(\bx-(\bn\cdot\bx)\bn)}{1+\alpha^2+2\alpha(\bn\cdot\bx)},\eeq
and therefore \beq\begin{array}{rcl}
s_\bn(\bx')&=&\frac{\bx'-(\bn\cdot\bx')\bn}{1-(\bn\cdot\bx')}=\frac{(1-\alpha^2)(\bx-(\bn\cdot\bx)\bn)}{(1-\alpha)^2+(1-\alpha^2)(\bn\cdot\bx)}\\
&=&\frac{1-\alpha^2}{(1-\alpha)^2}\,\frac{\bx-(\bn\cdot\bx)\bn}{1-(\bn\cdot\bx)}=\frac{1+\alpha}{1-\alpha}\,s_n(\bx).\end{array}\eeq
Now, since $\alpha=\tanh(\eta/2),$ we have
$$\frac{1+\alpha}{1-\alpha}=\frac{\cosh(\eta/2)+\sin(\eta/2)}{\cosh(\eta/2)-\sinh(\eta/2)}=\frac{2\exp(\eta/2)}{2\exp(-\eta/2)}=\exp(\eta),$$
and therefore \beq s_\bn(\bx')=e^\eta s_\bn(\bx),\eeq so that the
family of M\"obius transformations $g_\bn(\epsilon),$ when
parametrized by $\eta=2\,\mbox{arctanh}(\alpha),$ act as a
one--parameter group of uniform dilations on the stereographic
projection $s_\bn (S^n)=\BR^n.$
\section{Iterated function systems of conformal maps\label{sec:ifs}}
Let $S$ be a set, let $\{w_i:\, i=1,2,\,\ldots\, ,N\}$ be a family
of maps $w_i:\, S\longrightarrow S,$ and let $p_i(x),\,
i=1,2,\,\ldots\, ,N$ be positive functions on $S$ satisfying
$\sum_{i=1}^N p_i(x)=1,\, \forall x\in S.$ The maps $w_i$ and the
functions $p_i(x)$ define what is called an {\em iterated function
system (IFS) with place dependent probabilities}\, - cf.
\cite{barnsley}. Starting with an initial point $x_0$ we select one
of the transformations $w_i$ with the probability distribution
$p_i(x_0).$ If $w_{i_1}$ is selected, we get the next point
$x_1=w_{i_1}(x_0),$ and we repeat the process again, selecting the
next transformation $w_{i_2},$ according to the probability
distribution $p_i(x_1).$ By iterating the process we produce a
random sequence of integers $i_0,i_1,\,\ldots\, $ and a random
sequence of points $x_k=w_{i_k}(x_{k-1})\in S,\, k=1,2,\,\ldots\, .$
In interesting cases the sequence $x_k$ accumulates on an "attractor
set" which has fractal properties. Instead of looking at the points
of $S$ we can take a dual look at the functions on $S.$ Let ${\cal
F}(S)$ be the set of all real--valued functions on $S.$ ${\cal
F}(S)$ is a vector space, and each transformation $w:\, S\rightarrow
S$ induces a {\em linear}\, transformation $w^\star: {\cal
F}(S)\rightarrow {\cal F}(S)$ defined by $(w^\star f)(x)=f(w(x)),\,
x\in S,\, f\in {\cal F}(S).$
\subsection{Markov operator\label{subsec:markov}}
Given an iterated function system $\{w_i,\, p_i(\, .\, )\}$ on $S$
one naturally associates with it a linear {\em Markov operator}
(sometimes called also the {\em transfer}\, operator)\, $T^*: {\cal
F}(S)\rightarrow {\cal F}(S)$ defined by \beq (T^*f)(x)=\sum_{i=1}^N
p_i(x)(w_i^*f)(x)=\sum_{i=1}^N p_i(x)f(w_i(x)).\eeq There is a dual
Markov operator $T_*,$ acting on measures on $S.$ Suppose $S$ has a
measurable structure, $w_i$ and $p_i(\, .\, )$ are measurable, and
let ${\cal F}(S)$ be the space of all bounded measurable functions
on $S.$ Let ${\cal M}(S)$ be the space of all finite measures on
$S.$ Then $T_*: {\cal M}(S)\rightarrow {\cal M}(S)$ is defined by
duality: $(T_*\mu,f)=(\mu ,T^* f),$ where $(\mu ,f)\doteq \int fd\mu
.$ Since $T^*(1)=1,$ where $1(x)=1,\, \forall x\in S,$ we have that
$\int dT_*\mu=\int d\mu$ and, in particular, $T_*$ maps
probabilistic measures into probabilistic measures. In many
interesting cases the sequence of iterates $(T_*)^k\mu$ converges,
in some appropriate topology, to a limit
$\mu_\infty=\lim_{k\rightarrow\infty} (T_*)^k\mu ,$ that is
independent of the initial measure $\mu,$ and which is the unique
fixed point of $T_*.$ The support set of $\mu_\infty$ is then the
attractor set mentioned above.

Let $\mu_0$ be a fixed, normalized measure on $S,$ and assume that
the maps $w_i^{-1}$ map sets of measure $\mu$ zero into sets of
measure $\mu$ zero. Then, for any finite $k$, the measure
${T_\star}^k\mu_0$ is continuous with respect to $\mu_0$ and
therefore can be written as \beq {T_\star}^k\mu_0
(\br)=f_k(\br)\,\mu_0(\br).\label{eq:tmu0}\eeq The sequence of
functions $f_k(\br )$ gives a convenient graphic representation of
the limit invariant measure. In our case, as it follows from the
formula (\ref{eq:tmu0}), the maps $w_i$ are bijections, and the
functions $f_k$ can be computed explicitly via the following
recurrence formula: \beq f_{k+1}(\br )=
\sum_{i=1}^{N}p_i\left(w_i^{-1}(\br
)\right)\frac{d\mu_0\left(w_i^{-1}(\br)\right)}{d\mu_0(\br
)}\,f_k\left(w_i^{-1}(\br)\right). \label{eq:rec1}\eeq
\subsection{Conformal maps\label{subsec:cm}}In this section the set $S$ is either the
sphere $S^n,$ or the closed ball ${\bar B}^{n+1},$ and the maps $w$
are of the form (\ref{eq:gxg1}), and are determined by vectors
$\alpha\bn\in B^{(n+1)}.$ Let us choose one $\alpha,$ $0<\alpha <1,$
and $N$ unit vectors $\bn_i\in S^{n},$ so that we have $N$ maps \beq
w_i(\bx )=\frac{(1-\alpha^2)\bx+2\alpha(1+\alpha(\bn_i\cdot\bx
))\bn_i}{1+\alpha^2+2\alpha(\bn_i\cdot\bx)},\label{eq:wi}\eeq as in
Proposition \ref{prop:4}. In our case we have an additional
structure in the set $S$ and in the maps $w_i,$ namely the one
stemming from the Clifford algebra realization. First of all to each
$\bx\in S^n$ we have associated the idempotent $\frac{1}{2}P(\bx),$
where $P(\bx)=(1+\bx),$ and then we have a special class of
functions on $S,$ namely the functions of the form: \beq
f_a(\bx)=(P(\bx),a),\, a\in \cc,\, \bx\in {\bar
B}{(n+1)}.\label{eq:fa}\eeq We denote by ${\cal L}$ the vector space
of these functions. Notice that functions in ${\cal L}$ separate the
points $\bx\in {\bar B}^{(n+1)}.$ Indeed,  for $\bx ,\by \in {\bar
B}^{(n+1)}$ we have $f_{\by}(\bx )=\bx\cdot\by/2,$ thus our
statement reduces to: for any two different vectors $\bx_1,\bx_2$
one can always find another vector $\by$ such that
$\bx_1\cdot\by\neq \bx_2\cdot\by ,$ which is evident. \footnote{The
space ${\cal L}$ is $(n+2)$--dimensional, as it is clear that
$f_a(\bx )= 0,\, \forall\, a\in C_p\subset \cc,\ p> 1.$}
\begin{proposition}\label{prop:5}
With the notation as in the beginning of this section, let
$0<\alpha<1,$ $\bn_i\in S^n,\, i=1,2,\,\ldots\, N$ and $w_i$ as in
(\ref{eq:wi}). Suppose that
\begin{enumerate}
\item[{\rm 1)}] \beq\sum_{i=1}^N \bn_i=0,\label{eq:p11}\eeq
\item[{\rm 2)}] \beq p_i(\bx)=\frac{1+\alpha^2+2\alpha (\bn_i\cdot\bx)
}{Z(\alpha)},\label{eq:p12}\eeq where $$Z(\alpha)=\sum_{i=1}^N
(1+\alpha^2+2\alpha(\bn_i\cdot\bx) )=N(1+\alpha^2),$$
\end{enumerate}
then the Markov operator $T^*$ of the iterated function system
$\{(w_i,p_i)\}$ maps the space ${\cal L}$ into itself: $T^*:
f_a\mapsto f_V(a),$ where \beq
V(a)=\frac{1}{N(1+\alpha^2)}\sum_{i=1}^N P(\alpha\bn_i)\,
a\,P(\alpha\bn_i ).\label{eq:va}\eeq
\end{proposition}
\noindent {\bf Proof}: From (\ref{eq:gxg}) it follows that if
$\sum_i \bn_i =0,$ then $Z\doteq \sum_i
(1+\alpha^2+2\alpha(\bn_i\cdot\bx)
 )=N(1+\alpha^2)/(1-\alpha^2)$ is a constant, independent of $\bx.$
From the very definition of the Markov operator, as well as from
(\ref{eq:fa}),(\ref{eq:qaq}) it follows then that
\begin{eqnarray*}(T^*f_a)(\bx )&=&\sum\limits_i
p_i(\bx )f_a(w_i(\bx))=\sum_i p_i(\bx)\Phi(a\,P(w_i(\bx
)))\\
&=&\sum\limits_i
p_i(\bx)\Phi\left(a\frac{1-\alpha^2}{(1+\alpha^2+2\alpha(\bn_i\cdot\bx))}P(\alpha\bn_i)P(\bx)P(\alpha\bn_i)\right)\\
&=&\sum\limits_ip_i(\bx)\,\frac{(1-\alpha^2)}{1+\alpha^2+2\alpha(\bn_i\cdot\bx) }\Phi\left(P(\alpha\bn_i)aP(\alpha\bn_i)P(\bx)\right)\\
&=&\frac{1}{Z(\alpha)}\sum\limits_i\Phi\left(P(\alpha\bn_i)aP(\alpha\bn_i)P(\bx)\right)=f_{V(a)}(\bx).\\
\end{eqnarray*}\qed

The Markov operator $T^*$ acts on measures, while its dual $T^*$
acts on functions on $S.$ Every probabilistic measure $\mu$ on $S$
determines an algebra element $P(\mu )$ defined by: \beq P(\mu
)=\int\limits_S P(\bx)\, d\mu (\bx)=1+\int\limits_S \bx\, d\mu
(\bx)=P\left(\int\limits_S\bx\,d\mu(\bx )\right),\eeq so that
automatically $\Phi(P(\mu ))=1.$ $P(\mu )/2$ is an idempotent if and
only if $\mu$ is concentrated at just one point on the boundary
$S^n.$ In general there are infinitely many measures $\mu$ giving
rise to the same algebra element $P(\mu ).$ The process of
integration on one hand leads to simplification (linearization) but,
on the other hand, it also leads to the loss of information.
\begin{corollary}
Under the assumptions 1) and 2) of Proposition \ref{prop:5}, if
$\mu_1$ and $\mu_2$ are two probabilistic measures on $S$ such that
$P(\mu_1)=P(\mu_2)=P,$ then $P(T^*\mu_1)=P(T^*\mu_2)=V(P),$ where
$V(P)$ is given by the formula (\ref{eq:va}), with $a$ replaced by
$P.$\label{cor:2}
\end{corollary}
\noindent {\bf Proof}: Because functions $f_a,\, a\in \cc$ separate
the elements of $\cc,$ it is enough to show that
$f_a(P(T^*\mu))=f_a(V(P(\mu)))$ for all $a\in \cc.$ Now, from the
very definition of the functions $f_a,$ $f_a(\bx)=\Phi(aP(\bx )),$
and from the linearity of the trace functional $\Phi,$ it follows
that $(f_a,\mu)\doteq \int f_a(\bx )d\mu(x) = \Phi (aP(\mu)),$ and
so $f_a(V(P(\mu ))=\Phi(aV(P(\mu)))=\Phi(V(a)P(\mu))=f_V(a)(P(\mu
))=f_a(P(T^*\mu)).$ \qed \vgap
\begin{corollary}
Under the assumptions 1) and 2) of Proposition \ref{prop:5}, the
Markov operator recurrence formula (\ref{eq:rec1}) is explicitly
given by \beq
f_{k+1}(\br)=\frac{\phantom{N}(1-\alpha^2)^{n+2}}{N(1+\alpha^2)}\;
{\sum_{i=1}^N}\;\;
\frac{f_k\left(w_i^{-1}(\br)\right)}{\left(1+\alpha^2-2\alpha(\bn_i\cdot\bx)\right)^{n+1}}
,\label{eq:fn}\eeq where \beq
w_i^{-1}(\br)=\frac{(1-\alpha^2)\br-2\alpha(1-\alpha(\bn_i\cdot\br
))\bn_i}{1+\alpha^2-2\alpha(\bn_i\cdot\br)}.\eeq\label{cor:3}
\end{corollary}
\noindent {\bf Proof}: Follows directly by a somewhat lengthy
calculation using (\ref{eq:rec1}), (\ref{eq:wi}), (\ref{eq:p12}),
and (\ref{eq:dV}). \qed \begin{remark} Iterated function systems for
mixed states have been discussed by {\L}ozinski et al. in Ref.
\cite{loslzy}, while S{\l}omczynski \cite{slomczynski} discussed
Markov operators and dynamical entropy for general IFS--s on state
spaces. In these references the probability distributions assigned
to the maps were generic rather than derived geometrically, as is
the case in this paper.\end{remark}
\section{Examples\label{sec:examples}}
\subsection{$S^1$ -- Polygon\label{subsec:s1}} 
\begin{figure}[h!]
\begin{center}
    \leavevmode
      \includegraphics[width=10cm, keepaspectratio=true]{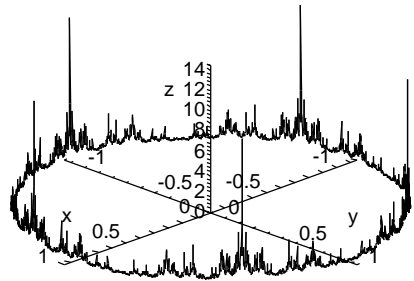}
\end{center}
  \caption{Pentagon. 7--th power of the Markov operator applied to $f= 1$.}\label{fig:pentagon}
\end{figure}

As the first example we consider the
circle $S^1,$ and unit vectors $\bn_i$ pointing to the vertices of a
regular polygon. For an illustration we choose the pentagon. Fig.
\ref{fig:pentagon} shows the plot of $\log_{10}(f_7+1.0),$ the 7--th
iteration of the Markov operator -- see (\ref{eq:fn}), for
$\alpha=0.58.$

\subsection{$S^2$\label{subsec:s2}}
\begin{figure}[h!]
\begin{center}
    \leavevmode
      \includegraphics[width=10cm, keepaspectratio=true]{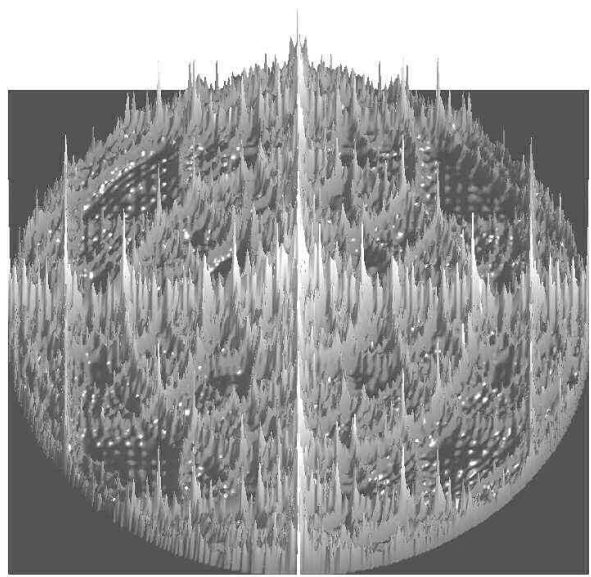}
\end{center}
  \caption{Octahedron -- \{3,4\}. 7--th power of the Markov operator, $\alpha=0.5.$}\label{fig:octahedron}
\end{figure}

$S^2,$ the Riemann sphere, is the same as the complex projective
line $P^1(\BC)$ - the space of pure quantum states of the simplest
non--trivial quantum system, namely spin $1/2.$ Examples of quantum
fractals on $S^2,$ based on Platonic solids, has been given
elsewhere (cf. \cite{jad06b}, and references therein). Here we give
just one example, namely the octahedral quantum fractal. Fig.
\ref{fig:octahedron} shows the 7--th power of the Markov operator:
$\log_{10}(f_7+1),$ - cf. (\ref{eq:fn}) for $\alpha=0.5,$ plotted on
the projection of the upper hemisphere of $S^2.$ The emergence of
circles on the plot is rather surprising and not well
understood.\footnote{The algorithm for generating conformal quantum
fractals on $S^2$ has been included in the CLUCalc software by
Christian Perwass. A video zooming on a quantum fractal based on the
regular octahedron, $\alpha=0.42,$  can be seen on the CLUCalc home
page: {{\protect \tt http://www.clucalc.info/}} }
\subsection{$S^3$ -- regular polytopes\label{subsec:s3a}}
There are six regular polytopes in four dimensions: self--dual
pentachoron (or $4$ simplex), $16$ cell (or cross--polytope, or
hexadecochoron), dual to it $8$ cell (or hypercube or tesseract),
self--dual $24$ cell (or icositetrachoron), $600$ cell (or
hexacosichoron), and its dual $120$ cell (or hecatonicosachoron) -
cf. Fig. 3 and Fig. 8. In our examples of four dimensional quantum
fractals we skip the first one. The pentachoron (the four
dimensional equivalent of the tetrahedron) leads to rather trivial
and uninteresting fractal pattern.
\subsection{$S^3$ -- $16$ cell.\label{subsec:s3b}}
Quaternions of the form $a+b{\bf i}+c{\bf j}+d{\bf k},\quad
a,b,c,d\in\BZ$ form the so called {\em Lipschitz\,} ring. The unit
quaternions of this ring form a group of order $8$ - the binary
dihedral group $D4.$ Its eight elements, $\{\pm 1,\pm{\bf i},\pm
{\bf j},\pm {\bf k}\}$ form the four-dimensional regular polytope,
the so called {\em cross--polytope\,}, with Schl\"afli symbol
$\{3,3,4\}.$ It has $16$ tetrahedral cells, each of its $24$ edges
belongs to $4$ cells.

Visualization of quantum fractals that live in four dimensions is
difficult. Here we generate 10,000,000 points of the iterated
function system described in Sec \ref{subsec:cm}, with $\bn_i$ being
the 8 vertices of the 16 cell, $\alpha=0.5,$ and with probabilities
given by (\ref{eq:p12}). We plot the three dimensional projections
of those 16742 points which fall into the slice of $S^3$ with the
fourth coordinate $0.5<x^4<0.51$ - see Fig. 4.

This pattern, generated by the IFS of conformal maps with
place-\-dependent probabilities should be compared with the plot of
the fourth approximation to the density of the limiting invariant
measure - Fig. 4. Due to the recursive nature of the formula
(\ref{eq:fn}) the computation time of $f_k$ grows exponentially with
$k.$ With each level new details appear in the graph, at the same
time the probability peaks get higher (as in Fig. 6). To present
more details in the graph, we are plotting $\log_{10} (f_4(\br)+1),$
rather than the function $f_4(\br)$ itself. Notice that for each
$k,$ the integral of $f_k(\br)$ over the sphere $S^3$, with the
natural $SO(4)$ invariant measure , is constant and equal to the
volume of $S^3.$

\subsection{$S^3$ -- $8$ cell.\label{subsec:s3c}}
\begin{figure}[h!]
\begin{center}
    \leavevmode
      \includegraphics[width=11cm, keepaspectratio=true]{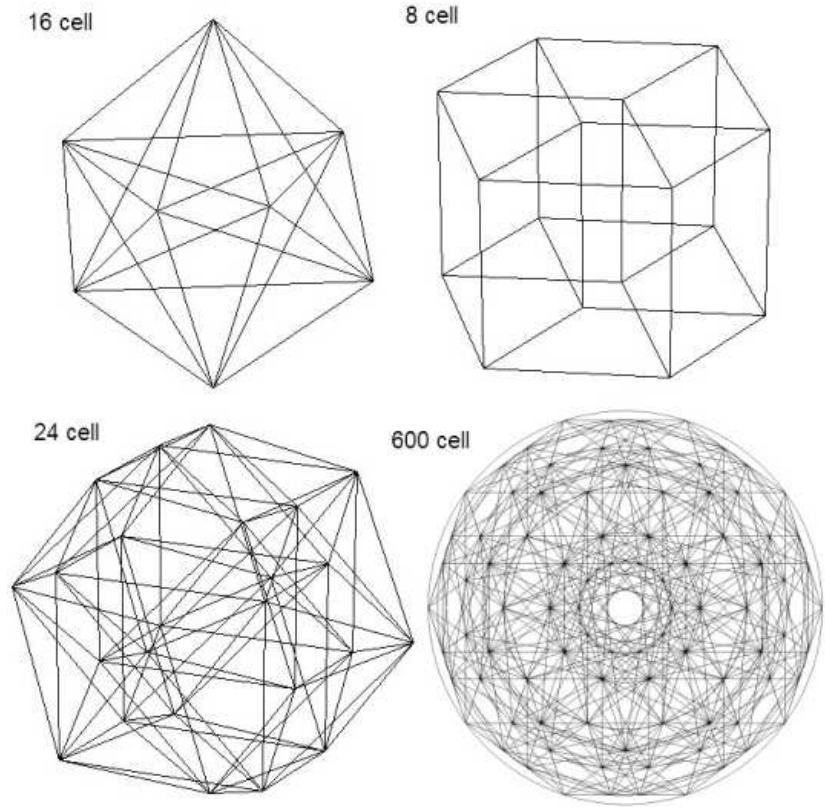}
\end{center}
  \caption{a) $16$ cell - \{3,3,3\}. $8$ vertices, $24$ edges, $32$ triangular faces, $16$ tetrahedral cells.
  b) $8$ cell or Hypercube - \{4,3,3\}.
  $16$ vertices, $32$ edges, $24$ square faces, $8$ cubic cells.
  c) $24$ cell - \{3,4,3\}. $24$ vertices, $96$ edges, $96$ triangular
  faces, $24$ octahedral cells.
  d) $600$ cell - \{3,3,5\}. $120$ vertices, $720$ edges, $1200$ triangular faces, $600$
  tetrahedral cells.  The graphics was generated by choosing the tetrahedral cell with vertices $t_0=(1,0,0,0),$ $t_1=(\phi,\Phi,0,1)/2,$ $t_2=(\phi,0,1,\Phi)/2,$ $t_3=(\phi,1,\Phi,0)/2,$ and choosing the unit vector
  $f_1$ in the direction of the center of this cell $(t_0+t_1+t_2+t_3)/4.$ The second unit vector $f_1$ was chosen in the direction of $f_0*t_1,$ (the quaternionic product). Then the frame
  $(f_0,f_1,f_2=(0,0,1,0),f_3=(0,0,0,1))$ was orthonormalized to $(e_0,e_1,e_2,e_3)$ via Gram-Schmidt procedure, and the $720$ edges of the $600$ cell have been projected onto $(e_2,e_3)$ plane.\label{fig:cellall}}
\end{figure}

Dual to the 16 cell is the $8$ cell, also known as {\em cross
polyhedron\,} {\em hypercube\,}, or {\em tesseract\,}. Its $16$
vertices are the unit quaternions $\frac{1}{2}(\pm 1,\pm {\bf i},\pm
{\bf j},\pm {\bf k}).$ Its Schl\"afli symbol is $\{4,3,3\},$ which
means that its cells are $\{4,3\}$ - that is cubes, each face
belongs to $2$ cells, and each edge belongs to $3$ cells. The
hypercube is built of two 3 dimensional cubes, their edges being
connected along the fourth coordinate. The projection of the
hypercube is shown in Fig. \ref{fig:cellall}.

We choose $16$ unit vectors $\bn_i$ pointing to the vertices of the
hypercube. Fig. \ref{fig:hcm} shows the plot of $f_5,$ the 5--th
iteration of the Markov operator (given by (\ref{eq:fn}), for
$\alpha=0.60,$ restricted to the section $x^3=0.8,$ projected onto
$(x^1,x^2)$ plane.

\subsection{$S^3$ -- 24 cell.\label{subsec:s3d}}
Quaternions of the form $a+b{\bf i}+c{\bf j}+d{\bf k},\quad
a,b,c,d\in\BZ$ or $a,b,c,d\in\BZ+\frac{1}{2}$ form a ring, called
the {\em Hurwitz ring\,}. Its additive group is the $F_4$ lattice.
The unite quaternions of this ring form a group, the {\em binary
tetrahedral group\,} $T_24,$ isomorphic to the group $SL(2,3)$ -
with generators the same as for $SL(2,5),$ - cf. (\ref{eq:sl25}),
except that the multiplications are carried in $\BZ_3.$ 24 cell has
Schl\"afli symbol $\{3,4,3\},$ which means that its 24 cells are
octahedrons, with each edge belonging to three cells \cite[p.
68]{duval}. Each of its 16 vertices is common to 6 cells - cf. Fig.
\ref{fig:cellall}. Fig. 6 shows the plots of $\log(f_k+1)$ for
$k=2,3,4,$ for $x^4=0.5,$ and $\alpha=0.6.$ With each power of the
Markov operator more details of the limit measure appear.

\subsection{$S^3$ -- $600$ cell.\label{subsec:s3e}}
Here we provide an example of a quantum fractal on $S^3,$ based on
the regular polytope in four dimensions, namely the $600$ cell, with
Schl\"afli symbol $\{3,3,5\}.$ The vertices of the $600$ cell are
given in the Appendix \ref{sec:big}. (c.f. also \cite[p.
74--75]{duval}.) Fig. 3 shows a two dimensional projection of the
600 cell as viewed from the direction of the center of one of its
cells, while Fig. 7 (top) shows the more perfect (all 120 vertices
can bee seen) Coxeter's projection. The inner ring, consisting of 30
vertices is on the torus. We show the functions $\log_{10}(f_1+1)$
and $log_10(f_2+1)$ plotted at the surface of this torus. The 30
highest peaks that can be seen on the bottom plots are located at
the vertices.

\subsection{$S^3$ -- $120$ cell.}
The last example is the 120 cell, with 600 vertices. Fig. 8 (top)
shows a particular projection of this polytope, with one of its 120
octahedral cells plotted in bold. Below is the plot of
$\log_{10}(f_2+1),$ for $\alpha=0.9,$ at the upper hemisphere
circumscribing this cell.

\section{Summary and conclusions}
In the standard formulation of the quantum theory the imaginary unit
$i$ plays an important yet somewhat mysterious role: it appears in
front of the Planck constant $\hbar,$ and provides a one--to--one
formal correspondence between Hermitian ``observables" and
anti--Hermitian generators of one--parameter groups of unitary
transformations. In particular it is necessary in order to write the
time evolution equation for the wave function, with the energy
operator (the Hamiltonian) defining the evolution. But the imaginary
``i" is not needed for quantum jumps. In a theory where quantum
jumps are the driving force of the evolution, the real algebra
structure, with a real trace functional suffices. In the present
paper we have studied the simplest case of real Clifford algebras of
Euclidean spaces and demonstrated that from the algebra and from the
geometry a natural family of iterated function systems of conformal
maps leads to fractal structures and pattern formation on spheres
$S^n.$ In this way we open a way towards algebraic generalizations
of quantum theory that are based on discrete, algebraic structure,
as expressed in the late Einstein's vision quoted in the
Introduction.

Among the open problems we would like to point out particularly the
following ones.

\subsection{Existence and uniqueness of the invariant measure\label{subsec:im}}
While numerical simulations (see the next section), suggest that for
the class of iterated function systems discussed in this paper, the
attractor set and the invariant measure exists and is unique, we are
not able to provide a mathematical proof. Even if the spheres $S^n$
and balls $B^{n+1}$ are compact, the M\"obius transformations of
these spheres are non-contractive. The question of existence and
uniqueness of invariant measures for non-contractive iterated
function systems has been discussed in the mathematical literature
\cite{szarek1,szarek2,apanasov}, yet none of the sufficient
conditions seems to be easily applicable to our case. Apanasov has a
whole book devoted to conformal maps, yet we find that his criteria,
esp. Theorem 4.16 of Ref. \cite{apanasov},  are abstract and
difficult to apply. Therefore the problem of existence and
uniqueness of the invariant measure for IFS--s discussed in the
present paper remains open at this time.
\subsection{Fractal dimension as a function of the parameter
$\epsilon.$} Anticipating a positive answer to the above problem,
the next important question is the exact nature of the fractal
attractor as a function of the parameter $\epsilon.$ The numerical
simulations seems to suggest that the fractal dimension of the
attractor of our IFSs on $S^n$ decreases, starting from $n,$ for
$\epsilon=0.$ Yet our attempt to determine its behavior, even for
the simplest case of $S^1,$ met an obstacle. We tried to calculate
the correlation dimension for the pentagon case, described in
Example 1. To this end we generated $N=10,000,000$ points, using the
algorithm of Sec. \ref{sec:ifs}, and plotted, on the log--log scale
the function $C(N,r),$ where $r$ is the distance between two points,
and $C(r)$ is the relative number of pairs, out of $N$ points,
within this distance. More precisely, the correlation dimension $D$
is defined as \beq D=\lim_{r\rightarrow 0} \log
(C(r))/\log(r),\label{eq:cordim}\eeq where \beq
C(r)=\frac{1}{N^2}\lim_{N\rightarrow \infty} \sum_{i,j}^N
\Theta(|r-|x_i-x_j|),\label{eq:cor}\eeq $\Theta$ being the unit step
function. For the standard Cantor set the correlation dimension
algorithm gives the correct fractal dimension, namely $D=0.63\approx
\log(2)/\log(3).$ For the pentagon, with $\epsilon=0.58,$ (cf. Fig.
1) we get a reasonable straight line with the slope $D\approx 0.9,$
but with $\epsilon=0.925,$ when the expected fractal dimension
should be close to zero, we get a staircase. It is not clear whether
this is due to numerical artifacts, or is it a pointer towards the
possible multifractality of quantum fractals for high values of
$\epsilon.$
\section{Appendices}
\subsection{The boosts in $SO(1,n+1)$}
Let $e_\mu,\, \mu=0,1,2,\ldots\, ,\mu_{n+1},$ $e_0=1\in\BR,$ $e_i\in
V,\, i=1,2,\ldots\,n+1$ be an orthonormal basis in $V^1.$ Then the
two--fold covering homomorphism $\Lambda: \tg\rightarrow
SO^+(1,n+1),$ $g\mapsto \Lambda(g),$ can be written as $ge_\mu
g^\tau=\Lambda^\nu_{\phantom{\nu}\mu},$ or, more explicitly:\beq
ge_0g^\tau=\Lambda^0_{\phantom{0}0}e_0+\Lambda^i_{\phantom{0}}e_i,\quad
ge_ig^\tau=\Lambda^0_{\phantom{0}i}e_0+\Lambda^j_{\phantom{j}i}e_j.
\eeq If $x\in V^1$ is written in terms of the basis $e_\mu,$ $x=^\mu
e_\mu,$ then $x'=gxg^\tau=x'^\mu e_\mu$ is given by $x'^\mu
=\Lambda^\mu_{\phantom{\mu}\nu}x^\nu.$ It is then easy to see that
the map $\phi_g:\sn\rightarrow \sn,$ given by (\ref{eq:phi}), when
written in terms of the representing it matrix $\Lambda(m)\in
SO^+(1,n+1),$ is \beq
\phi_\Lambda(\bx)^i=x'^i/x'^0=\frac{\Lambda^i_{\phantom{i}0}+\Lambda^i_{\phantom{i}j}x^j}{\Lambda^0_{\phantom{0}0}+\Lambda^0_{\phantom{0}j}x^j},\quad
i,j=1,2,\ldots\, ,n+1\label{eq:philambda}\,\eeq where
$\bx^2=\sum_i^{n+1}(x^i)^2=1.$
\begin{proposition}
The map $\phi_\Lambda: \sn\longrightarrow \sn$ given by: \beq
\phi_\Lambda(\bx)^i=x'^i/x'^0=\frac{\Lambda^i_{\phantom{i}0}+\Lambda^i_{\phantom{i}j}x^j}{\Lambda^0_{\phantom{0}0}+\Lambda^0_{\phantom{0}j}x^j},\quad
i,j=1,2,3\label{eq:phil},\eeq transforms the normalized $\sn$
invariant measure $dS$ on $\sn$ into a new measure
$dS'=\phi_\Lambda^*(dS),$ where $\phi_\Lambda^*(dS)$ is the
pullback, (or the ``inverse image", cf. e.g. \cite[Ch.
16.20.8]{dieudonne}) of $dS$ by $\phi_\Lambda$). For $\bx\in \sn$ we
have \beq \left(\phi_\Lambda^*(dS)\right)(\bx)=
\frac{1}{\left(\Lambda^0_{\phantom{0}0}+\Lambda^0_{\phantom{0}i}\,\bx^i\right)^n}\,dS(\bx).\label{eq:dsds2}\eeq
\label{prop:6}\end{proposition}
\noindent To prove (\ref{eq:dsds2}) we will need a couple of lemmas.
\begin{lemma}
Let $r$ be a real number, and let $f_r:SO^+(1,n+1)\times
\sn\longrightarrow \BR$ be defined as \beq
f_r(\Lambda,\bx)=\left(\Lambda^0_{\phantom{0}0}+\Lambda^0_{\phantom{0}i}\bx^i\right)^r.\label{eq:coc}\eeq
Then $f_r$ has the following cocycle property: \beq
f_r(\Lambda\Lambda',\,\cdot\,)=\phi_{\Lambda'}^*\left(f_r(\Lambda,\,\cdot\,)\right)f_r(\Lambda',\,\cdot\,).\eeq
\label{lem:coc}\end{lemma}
\noindent {\bf Proof}: It is enough to consider the case $r=1.$ We
set, during the course of this proof, $f_1=f.$ We have
\begin{eqnarray*}f(\Lambda\Lambda',\bx)&=&(\Lambda\Lambda')^0_{\phantom{0}0}+(\Lambda\Lambda')^0{\phantom{0}i}\,\bx^i\\
&=&\Lambda^0_{\phantom{0}0}{\Lambda'}^0_{\phantom{0}0}+\Lambda^0_{\phantom{0}i}{\Lambda'}^i_{\phantom{i}0}+
\Lambda^0_{\phantom{0}0}{\Lambda'}^0_{\phantom{0}i}\bx^i+\Lambda^0_{\phantom{0}k}{\Lambda'}^k_{\phantom{k}i}\bx^i\\
&=&\Lambda^0_{\phantom{0}0}\left({\Lambda'}^0_{\phantom{0}0}+{\Lambda'}^0_{\phantom{0}i}\bx^i\right)+\Lambda^0_{\phantom{0}k}
\left({\Lambda'}^k_{\phantom{k}0}+{\Lambda'}^k_{\phantom{k}i}\bx^i\right)\\
&=&\left({\Lambda'}^0_{\phantom{0}0}+{\Lambda'}^0_{\phantom{0}i}\bx^i\right)\left(\Lambda^0_{\phantom{0}0}+
\Lambda^0_{\phantom{0}k}
\frac{{\Lambda'}^k_{\phantom{k}0}+{\Lambda'}^k_{\phantom{k}j}\bx^j}
{{\Lambda'}^0_{\phantom{0}0}+{\Lambda'}^0_{\phantom{0}j}\bx^j}\right)\\
&=&f(\Lambda',\bx)f(\Lambda,\phi_{\Lambda'}(\bx)).
\end{eqnarray*}
\qed\vgap
\begin{lemma}
Let $m\in\tgp,$ and let $\Lambda=\Lambda(m)\in SO^+(1,n+1)$ be the
matrix representing $m$. Then (\ref{eq:dsds2}) holds for $\Lambda.$
\label{lem:pos}\end{lemma}
\noindent{\bf Proof}: It is enough to consider the case of $m\neq
I.$ Let us write $m$ in the form $m=\frac{1}{\sqrt{1-\alpha^2}},$
$0<\alpha<1,$ $\bn^2=1,$ as in (\ref{eq:ai}). From (\ref{eq:gxg}),
and the general formula $gx^\mu e_\mu
g^\tau=x^\mu\Lambda(g)^\nu_{\phantom{\nu}\mu} e^\mu$ it follows that
$\Lambda^0_{\phantom{0}0}+\Lambda^0{\phantom{0}i}x^i$ is the
coefficient in front of $e_0$ on the right hand side of
(\ref{eq:gxg}), which is
$(1+\alpha^2+2\alpha(\bn\cdot\bx))/(1-\alpha^2).$ Comparing now
(\ref{eq:jac}) and (\ref{eq:dsds2}) we see that the two formulas
coincide.\qed\vgap
\noindent{\bf Proof of the Proposition \ref{prop:6}}: Let $g\in\tg$
and let $g^tau=mu$ be the decomposition of $g^\tau$ into a
spin--boost and a rotation as in (\ref{eq:dec}), so that $g=u\tau
m.$ Let $R=\Lambda(u^\tau).$ Since $R\in SO(n+1),$ and $dS$ is
rotation invariant, we have $\phi_R^*(dS)=dS.$ Notice that
(\ref{eq:dsds2}) can also be written as \beq
dS'=f_n(\Lambda(g),\,\cdot\, )\, dS.\label{eq:dsf2}\eeq Now, \beq
\phi_{\Lambda(g)}^*(dS)=\phi_{\Lambda(mu^\tau)}^*=\phi_{\Lambda(m)}(\phi_R^*(dS))=\phi_{\Lambda(m)}^*(dS)
=f_n(\Lambda(m),\,\cdot\,)\, dS,\eeq where we have used Lemma
\ref{lem:pos}.

\noindent Now, from the definition (\ref{eq:coc}) of the cocycle
$f_n$ we have that $f_n(R^{-1},\,\cdot\, )= 1,\, \forall\, R\in
SO(n+1).$ Therefore $$f_n(\Lambda(m),\,\cdot\,
)=f_n((R^{-1}R)\Lambda(m),\,\cdot\,
)=f_n(R^{-1}(R\Lambda(m)),\,\cdot\, )=f_n(R^{-1}\Lambda(g),\,\cdot\,
),$$ and from Lemma \ref{lem:coc}, and the rotational invariance of
$f_n$ mentioned before, we find $f_n(R^{-1}(\Lambda(g)),\,\cdot\,
)=\phi_{\Lambda(g)}^*(f_n(R^{-1},\,\cdot\, )f_n(\Lambda(g),\,\cdot\,
)=f_n(\Lambda(g),\,\cdot\, ).$ This proves that the formula
(\ref{eq:dsf2}) for a general $\Lambda\in SO^+(1,n+1),$ which is the
same as (\ref{eq:dsds2}). \qed\vgap

\subsection{Hamilton's Icosian
Calculus\label{sec:icosians}}
\begin{quotation}
Hamilton's ``Icosian Calculus'' dates back to his communication to
the Proc. Roy. Irish Acad. of November 10, 1856 \cite[p.609]{ham1},
followed by several papers, the last one in 1863. According to the
contemporary terminology Hamilton proposes a particular presentation
of the alternating group $A5$ - the symmetry group of the
icosahedron.
\begin{center}Account of the Icosian Calculus\\
Communicated 10 November 1856. \\
\[Proc. Roy. Irish Acad. vol. vi (1858), pp. 415-16.\]\end{center}

Sir William Rowan Hamilton read a Paper on a new System of Roots of
Unity, and of operations therewith connected: to which system of
symbols and operations, in consequence of the geometrical character
of some of their leading interpretations, he is disposed to give the
name of the ``ICOSIAN CALCULUS''. This Calculus agrees with that of
the Quaternions, in three important respects: namely,  1st that its
three chief symbols $\iota,\kappa,\lambda$ are (as above suggested)
roots of unity, as $i,j,k$  are certain fourth roots thereof: 2nd,
that these new roots obey the associative law of multiplication; and
3rd, that they are not subject to the commutative law, or that their
places as factors must not in general be altered in a product. And
it differs from the Quaternion Calculus, 1st, by involving roots
with different exponents; and 2nd by not requiring (so far as yet
appears) the distributive  property of multiplication. In fact, $+$
and $-$ , in these new calculations, enter only as connecting
exponents, and not as connecting terms: indeed, no terms, or in
other words, no polynomes,  nor even binomes, have hitherto
presented themselves, in these late researches of the author. As
regards the exponents of the new roots, it may be mentioned that in
the principal system - for the new Calculus involves a family of
systems-there are adopted the equations, $$ 1 = \iota^2 = \kappa^3 =
\lambda^5,\, \lambda = \iota\kappa ;\;\;\;\; (A)$$ so that we deal,
in it, with a new square root, cube root, and fifth root, of
positive unity; the latter root being the product of the two former,
when taken in the order assigned, but not in the opposite order.
From these simple assumptions (A), a long train of consistent
calculations opens itself out, for every result of which there is
found a corresponding geometrical interpretation, in the theory of
two of the celebrated solids of antiquity, alluded to with interest
by Plato in the Timaeus; namely the Icosahedron, and the
Dodecahedron: whereof the angles may now be unequal. By making
$\lambda^4 = 1,$ the author obtains other symbolical results, which
are interpreted by the Octahedron and the Hexahedron. The Pyramid
is, in this theory, almost too simple to be interesting: but it is
dealt with by the assumption, $\lambda^3 = 1,$ the other equations
(A) being untouched. As one fundamental result of those equations
(A), which may serve as a slight specimen of the rest, it is found
that if we make $\iota\kappa^2 = \mu,$ we shall have

$$ \mu^5=1,\, \mu = \lambda \iota \lambda,\, \lambda = \mu \iota
\mu;$$

\noindent so that this new fifth root mu  has relations of perfect
reciprocity with the former fifth root lambda. But there exist more
general results, including this, and others, on which Sir W. R. H.
hopes to be allowed to make a future communication to the Academy:
as also on some applications of the principles already stated, or
alluded to, which appear to be in some degree interesting.
\end{quotation}
\subsection{The Binary Icosahedral
Group\label{sec:big}} Putting $R=\iota ,$ $S=\kappa ,$
$T=\lambda^4,$ we can equivalently write Hamilton's equations $(A)$
(Sec. \ref{sec:icosians}) as \beq R^2=S^3=T^5=RST=1.\eeq Removing
the last equality we get the code for the {\em binary icosahedral
group:}\ \beq R^2=S^3=T^5=RST.\label{eq:rst}\eeq It is evident from
the definition that $Z=RST$ is a central element of the group, and
it can be shown \cite[p. 69 and references therein]{coxeter40c} that
$Z$ is of order 2: $Z^2=1.$ This group if order 120, denoted as
$2.A5,$ and it is a double cover of the icosahedral group $A5.$ The
group has a particularly simple representation in terms of the
quaternions. Let \beq \phi=\frac{1+\sqrt{5}}{2}=1.61803\ldots,\quad
\Phi=\frac{-1+\sqrt{5}}{2}=\phi^{-1}=0.61803\ldots,\eeq be the
Golden Ratio and its inverse, respectively. Consider the group $G$
consisting of $120$ elements given by Table 1 below:\\
\begin{table}[h]
\begin{center}
\caption{$120$ vertices of the $600$ cell} \label{600cell}
\vspace*{0.2in}
\begin{tabular}[nic]{|l|l|}
\hline $2\times 4=8$&elements of the form $(\pm
1,0,0,0),(0,\pm 1,0,0),$\\&$(0,0,\pm 1,0),(\pm 0,0,0,\pm 1)$\\
\hline
 $2^4=16$&elements of the form $(\pm \frac{1}{2},\pm
\frac{1}{2},\pm
\frac{1}{2},\pm\frac{1}{2})$\\
\hline
$3!\times 2^3=96$&elements that are even permutations of elements\\
&of
the form $\frac{1}{2}(\pm \phi,\pm 1,\pm \Phi,0).$\\
\hline
\end{tabular}
\end{center}
\end{table}

These $120$ elements form a group of unit icossians (cf. Appendix
\ref{sec:icosians}) that is a finite subgroup of the group
$Spin(3).$ For generators $R,S$ we can take, for
instance\footnote{One can check that there are $120$ possible
choices of triples of quaternionic generators $R,S,T$ satisfying
(\ref{eq:rst}).}, \beq S_1=\frac{1}{2}(1-\Phi i-\phi k),\quad
T_1=\frac{1}{2}(\Phi-i-\phi j),\quad R_1=S_1T_1=-i, \eeq or an
inequivalent set \beq S_2=\frac{1}{2}(1+\phi i+\Phi j),\quad
T_2=\frac{1}{2}(-\phi-i-\Phi k).\quad R_2=S_2T_2=-i. \eeq In both
cases we have $RST=-1,$ but the two sets of generators are
geometrically inequivalent (they are related by an {\em outer\,}
automorphism of $G$), the angle between $S_1$ and $T_1$ is $\pi/5$
while the angle between $S_2$ and $T_2$ is $3\pi/5.$

The binary icosahedral group is isomorphic to $SL(2,5),$ the group
of unimodular $2\times 2$ matrices over the field $Z_5,$ as can be
seen by taking for the generators $R,S,T$ the matrices: \beq
R=\begin{pmatrix}0&1\\-1&0\end{pmatrix},\,
S=\begin{pmatrix}1&-1\\1&0\end{pmatrix},\,
T=\begin{pmatrix}-1&0\\-1&-1\end{pmatrix}.\label{eq:sl25}\eeq

Fig. \ref{fig:cellall} shows the vertices of the 600 cell as viewed
from the direction of the center of one of its cells. There is
another realization of the 600 cell as a polytope, due to Coxeter
\cite[p.247]{cox63}, where all of the 120 vertices are organized on
four different tori within the sphere $S^3.$ Let
$$a=\sqrt{(1+3^{-1/2}5^{-1/4}\phi^{3/2})/2}\approx 0.947274,$$
$$b=\sqrt{(1+3^{-1/2}5^{-1/4}\phi^{-3/2})/2}\approx 0.770582,$$
$$c=\sqrt{(1-3^{-1/2}5^{-1/4}\phi^{-3/2})/2}\approx 0.637341,$$
$$d=\sqrt{(1-3^{-1/2}5^{-1/4}\phi^{3/2})/2}\approx 0.320426,$$
let $\theta=\pi/30,$ and let the four families, each of 30 vertices,
be given by: \beq \begin{array}{rclrr}a[k]&=& \{a\cos(k\theta),\,\,
a\sin(k\theta), &d\cos(11k\theta), &d\sin(11k\theta)\},\\
b[k]&=& \{d\cos(k\theta),\,\, d\sin(k\theta), &-a\cos(11k\theta),
&-a\sin(11k\theta)\},\\ \end{array}
 \eeq where $$k=0,k<60,k=k+2,$$ and
\beq \begin{array}{rclrr}a[k]&=& \{b\cos(k\theta),\,\,
b\sin(k\theta), &c\cos(11k\theta), &c\sin(11k\theta)\},\\
b[k]&=& \{c\cos(k\theta),\,\, c\sin(k\theta), &-b\cos(11k\theta),
&-b\sin(11k\theta)\},\\ \end{array}
 \eeq where $$k=1,k\leq 60,k=k+2.$$
\textbf{Acknowledgements}: Thanks are due to Pierre Angl\`es for his
criticism and invaluable advice, and to B. N. Apanasov, Robert P. C.
Marrais, Christian Perwass, F. D. (Tony) Smith (Jr), T. Szarek,
Jeffrey Weeks, K. Zyczkowski, as well as to Russell Towle and other
members of the GAP forum, for useful hints and pointers to the
literature. The author would also like to thank Cambridge University
Press for their kind permission to quote in extenso the paper by Sir
William Rowan Hamilton introducing the concept of the Icosian
calculus. Last, but not least, I thank my wife, Laura, for reading
the manuscript.
\newpage
\section{Figures\label{sec:fig}}
\newpage
\newpage
\begin{figure}[h!]
\begin{center}
    \leavevmode
      \includegraphics[width=7cm, keepaspectratio=true]{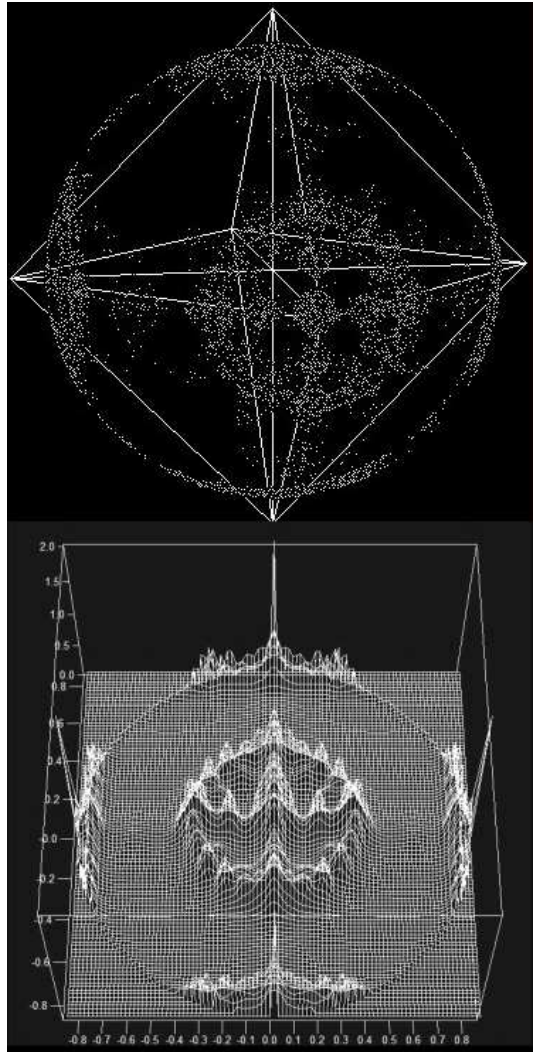}\label{fig:cell16rnd}
\end{center}
  \caption{$16$ cell -- \{3,3,4\}. Generated 10,000,000 random points of the IFS system of conformal maps with $\alpha=0.5.$
  Plotted are 16742 points whose fourth coordinate is in the slice $0.5<x^4<0.51.$ The picture is superimposed on the
  projection of the edges of the 16 cell. Below: Plotted the fourth power of the Markov operator, more precisely of the
  function $\log_{10}(f_4(\br)+1),$ with $f_4$
  function defined in (\ref{eq:fn}), calculated
  for $\alpha=0.5$ and $x^4=0.5.$ }
\end{figure}
\newpage
\begin{figure}[h!]
\begin{center}
    \leavevmode
      \includegraphics[width=10cm, keepaspectratio=true]{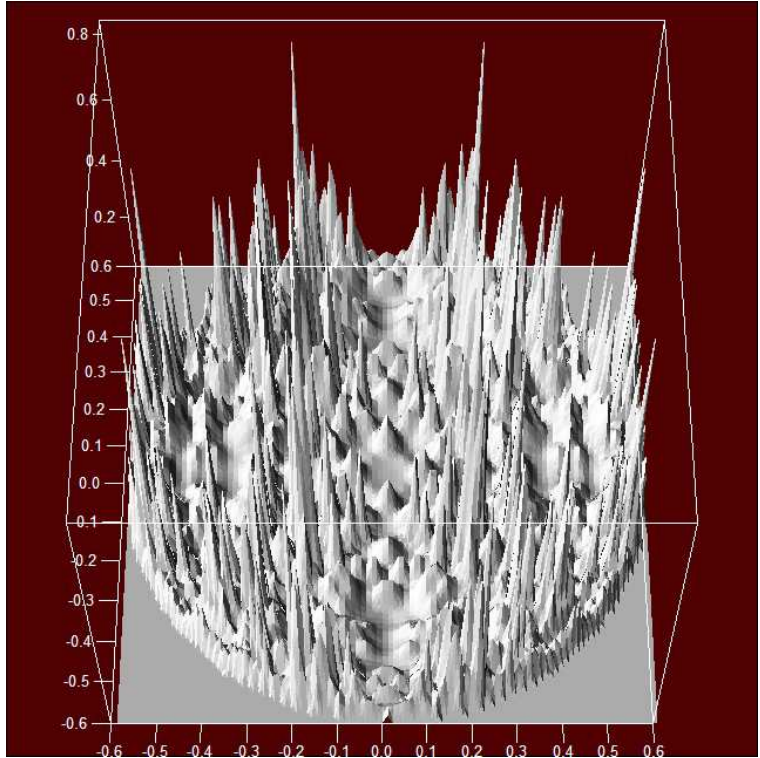}
\end{center}
  \caption{Hypercube -- \{4,3,3\}. $5$th power of the Markov operator, (\ref{eq:fn}), with $\alpha=0.6,$ computed
  at the section $x^4=0.8.$ Plotted is the $\log_{10}((f_5)+1).$}\label{fig:hcm}
\end{figure}
\newpage
\begin{figure}[h!]
\begin{center}
    \leavevmode
      \includegraphics[width=11cm, keepaspectratio=true]{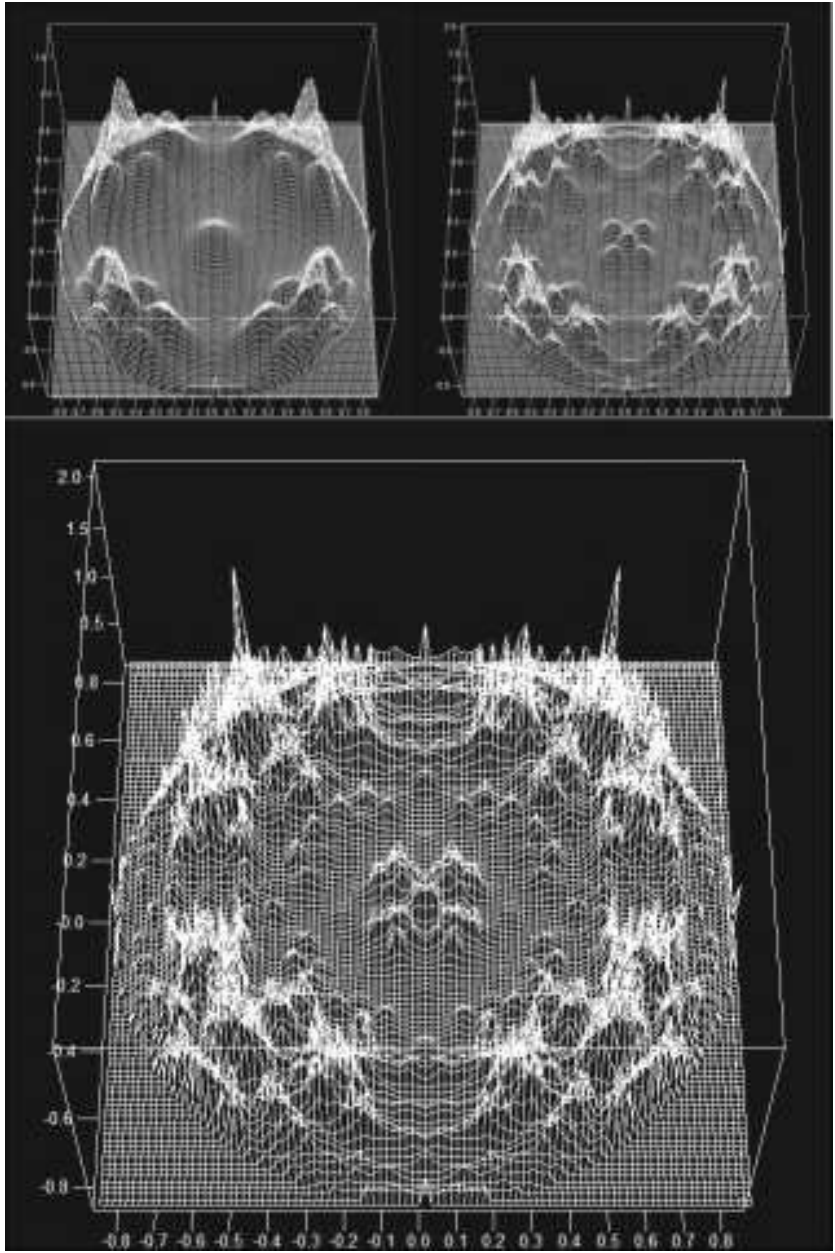}
\end{center}
  \caption{24 cell -- \{3,4,3\}. Markov operator levels 2,3 and 4, for $\alpha=0.6,$ plotted at $x^4=0.5.$}\label{fig:24cellml1}
\end{figure}
\newpage
\begin{tabular}{c}

     \includegraphics[width=8cm, keepaspectratio=true]{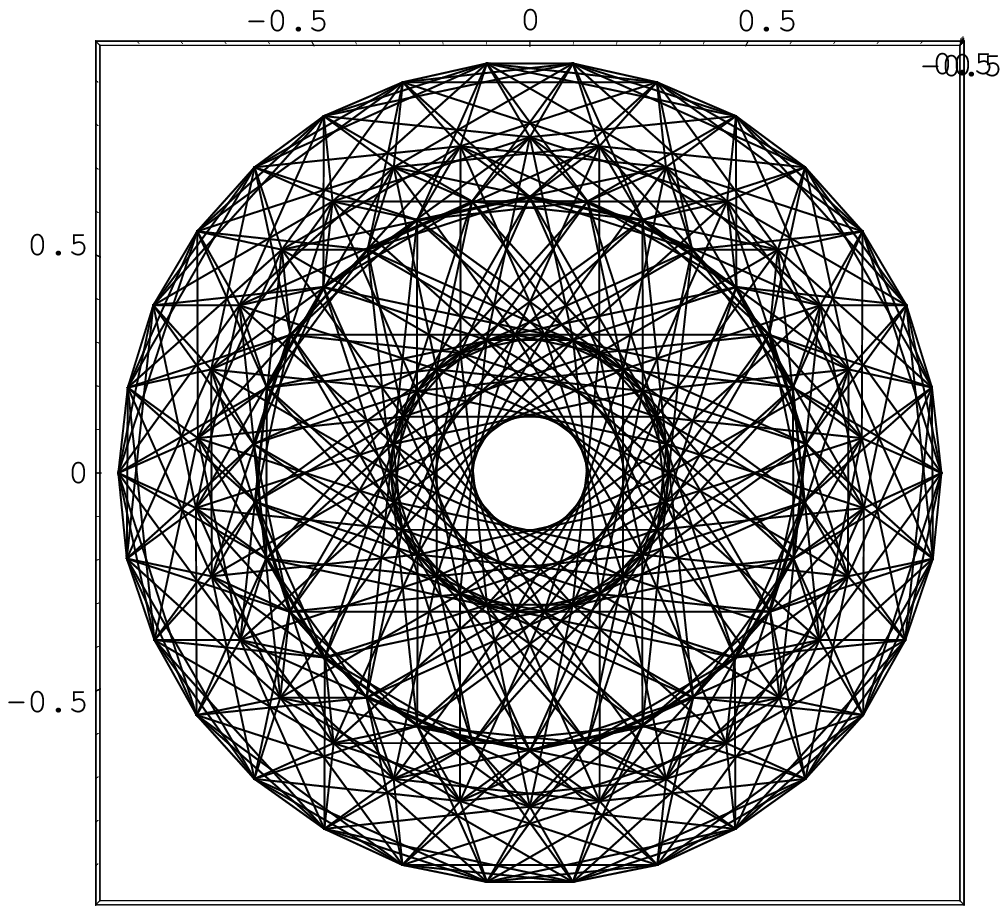}
\\

\includegraphics[width=8 cm, keepaspectratio=true]{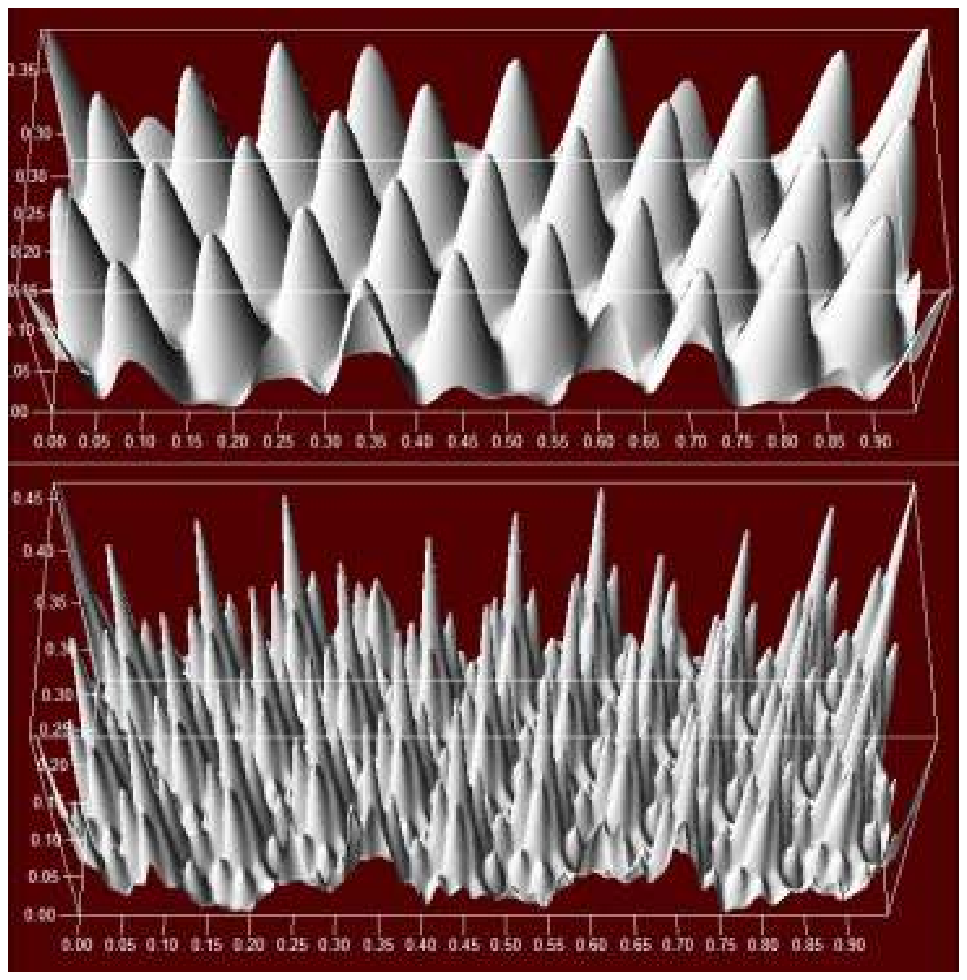}
\end{tabular}
\vgap\noindent Figure 7: 600 cell - \{3,3,5\}. Top: Coxeter's
projection. Below 1st and 2nd powers of the Markov operator, for
$\alpha=0.6$ plotted at the surface  of the most inner torus.

\newpage
\begin{tabular}{c}

     \includegraphics[width=8cm, keepaspectratio=true]{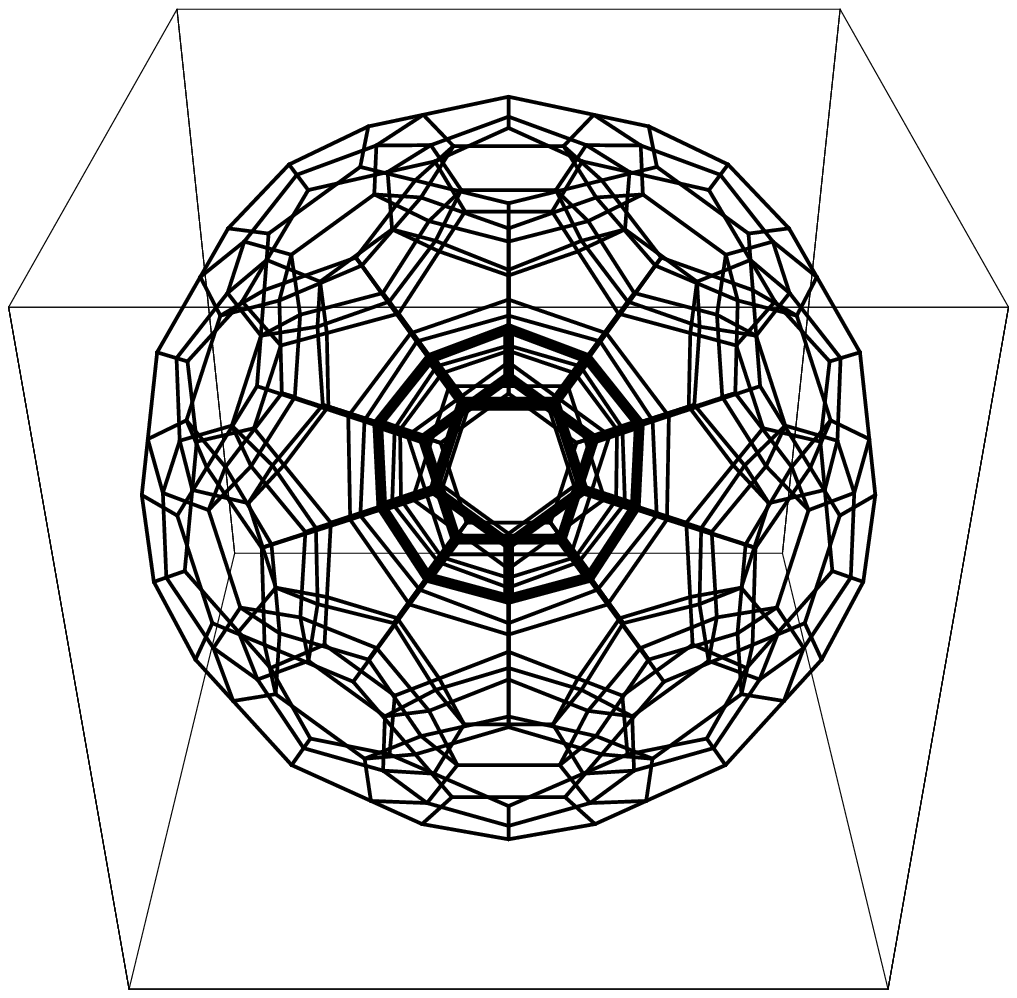}
\\

\includegraphics[width=8 cm, keepaspectratio=true]{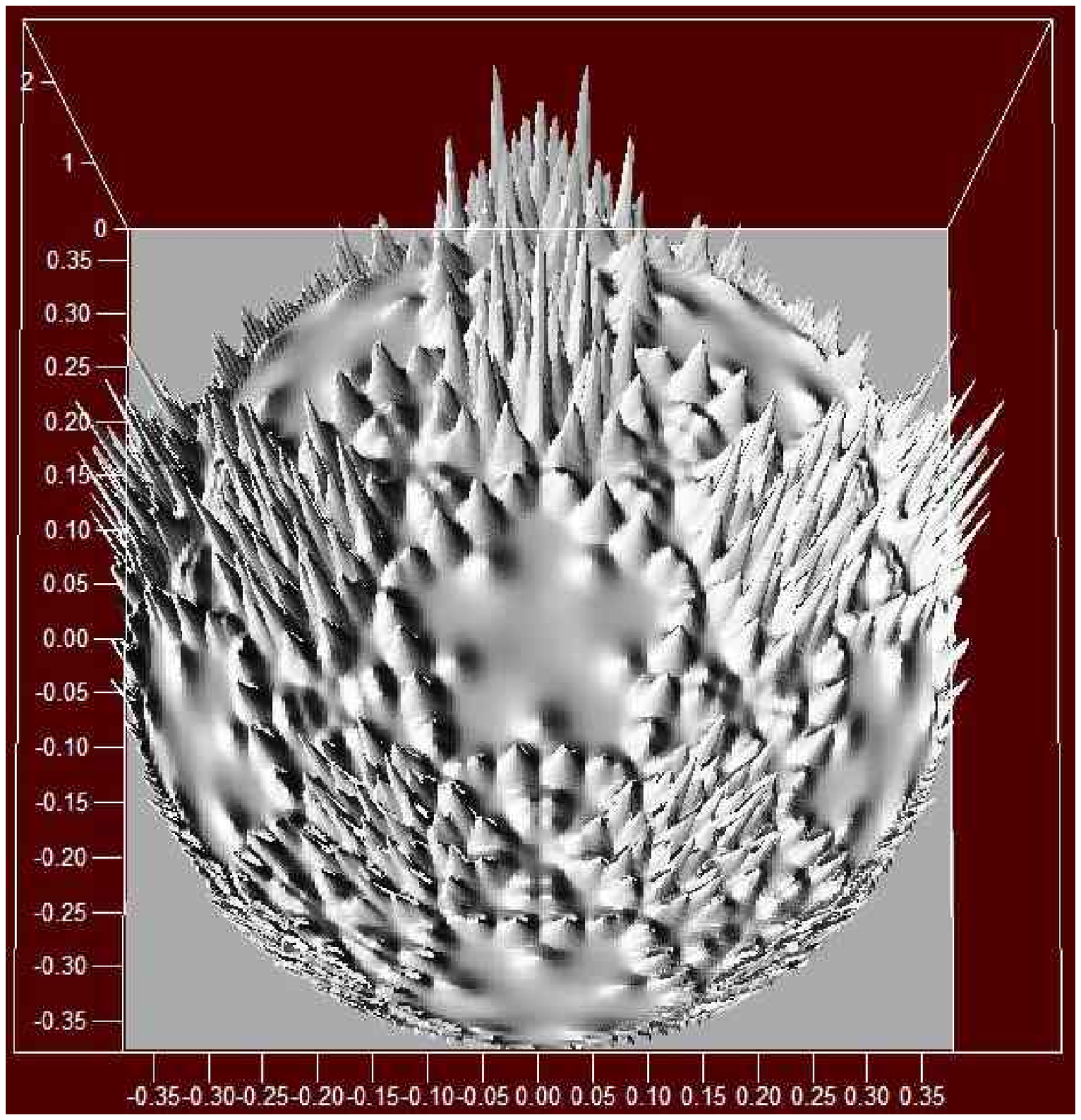}
\end{tabular}
\vgap\noindent Figure 8: 120 cell -- \{5,3,3\}. $600$ vertices,
$1200$ edges of length $(1-\phi)/\sqrt(2)$, $720$ pentagonal faces,
$120$ dodecahedral cells. One of its dodecahedral cells in bold.
Below the 2nd power of the Markov operator, for $\alpha=0.9,$
plotted at the upper hemisphere of this particular cell.

\begin{tabular}{c}

     \includegraphics[keepaspectratio=true]{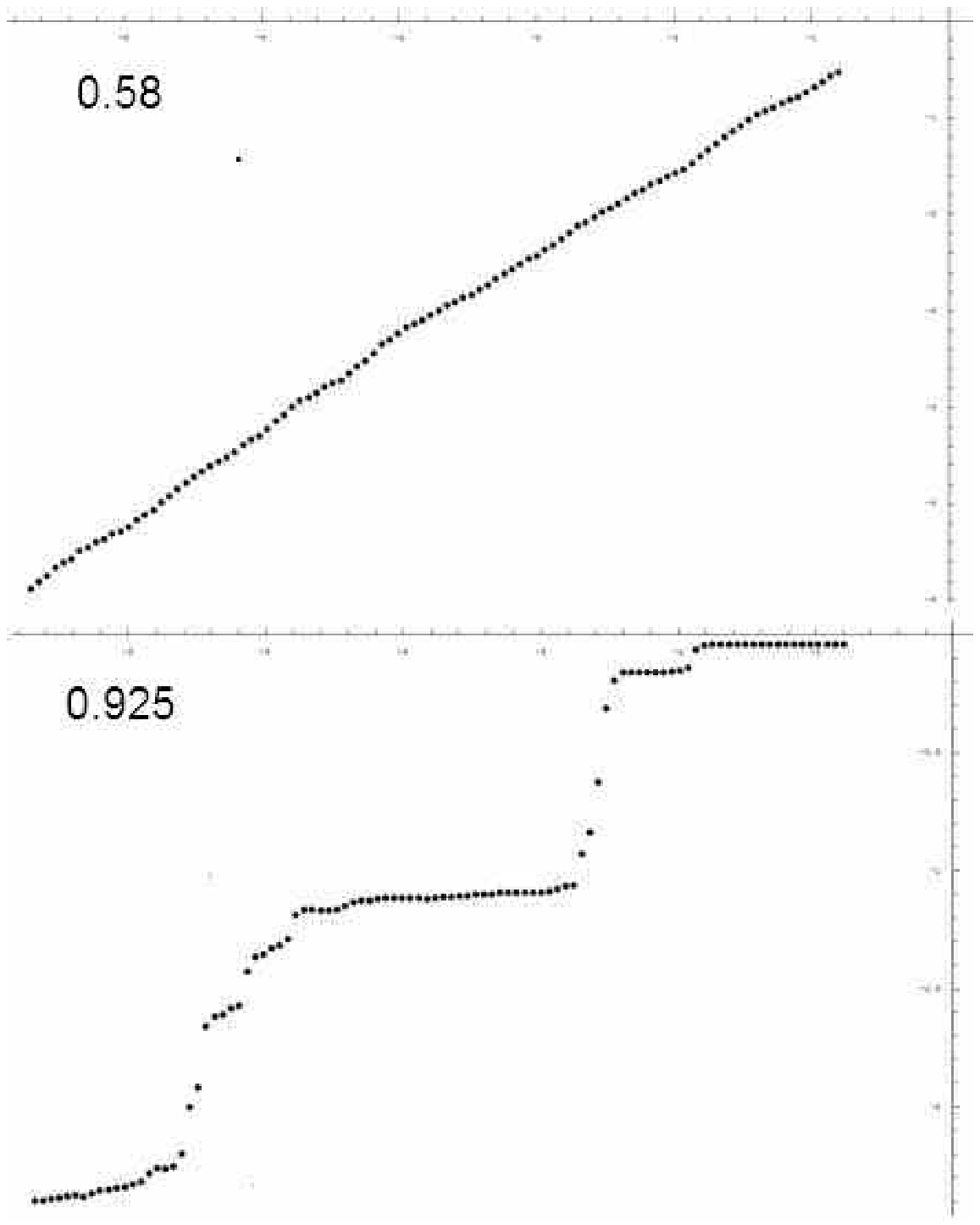}
\end{tabular}
\vgap\noindent Figure 9: Correlation dimension plots for the
pentagon, for $\alpha=0.58,$ and $\alpha=0.925.$ Plotted is the
function $\log(C(r)),$ defined in (\ref{eq:cor}), versus $\log(r).$
The slope of the graph should give the correlation dimension $D$ --
 (\ref{eq:cordim}). For $\epsilon=0.58,$ (cf. Fig. 1) we get a
reasonable straight line with the slope $D\approx 0.9,$ but with
$\epsilon=0.925,$ when the expected fractal dimension should be
close to zero, we get a staircase.

{\small \vskip 1pc {\obeylines }}
{\small \noindent Arkadiusz Jadczyk }\\ {\small \noindent IMP }\\
{\small \noindent France }\\
{\small \noindent E--mail: lark1@cict.fr }

\end{document}